\documentclass[showpacs,preprintnumbers]{revtex4}

\begin{document}

\title{Exotic baryons and the large-$N_{c}$ expansion}
\author{P.~V.~Pobylitsa}
\affiliation{Institute for Theoretical Physics II, Ruhr University Bochum, D-44780
Bochum, Germany\\
and\\
Petersburg Nuclear Physics Institute, Gatchina, St. Petersburg, 188300,
Russia}

\begin{abstract}
The status of exotic baryons is analyzed in the limit of a large number of
colors $N_c$. Several toy models reproducing the main features of the
large-$N_c$ QCD are studied in order to clarify the recent controversy concerning
the consistency of various approaches to the $1/N_{c}$-expansion for exotic
baryons. The analysis reveals a number of contradictions between the systematic
$1/N_c$-expansion and the rigid-rotator approximation, including the problems
of this approximation in the description of exotic baryons.
\end{abstract}

\maketitle

\section{Introduction}

The recent discovery \cite{Nakano,Barmin,Stepanyan,Barth,Asratyan} of the
$\Theta ^{+}$ baryon with a mass of 1540 MeV close to the values predicted
in Refs. \cite{Praszalowicz-87,DPP-97} has led to considerable excitement
in hadronic physics. One of the main ingredients involved in the prediction
of $\Theta ^{+}$ \cite{Praszalowicz-87,DPP-97} is the limit of the large
number of the QCD colors $N_{c}$ and the $1/N_{c}$-expansion. In spite of
the numerical agreement between the theoretical prediction of
Refs.~\cite{Praszalowicz-87,DPP-97} and the experimental data, recently certain doubts
have been expressed in Refs. \cite{Cohen-03a,Cohen-Lebed-03,CDNL-03,IKOR-03}
concerning the consistency of the methods used in
Refs.~\cite{Praszalowicz-87,DPP-97} with the systematic $1/N_{c}$-expansion.

As is well known, the standard picture of large-$N_{c}$ baryons assumes that
they can be described by a mean-field solution of some effective theory
equivalent to the large-$N_{c}$ QCD \cite{Witten-79}.
The precise action of this effective
theory is not known. Nevertheless, the general properties of large-$N_{c}$
baryons can be derived from the symmetry of the mean-field solution. It is
assumed that the mean-field solution violates both the flavor $SU(3)$ group
and the group $SO(3)\sim SU(2)$ of usual space rotations. However, the
mean field is supposed to be invariant under combined $SU(2)$ rotations in the
flavor and in the usual space. The realization of this symmetry can be found
in various models ranging from the naive quark model to the Skyrme model
\cite{Skyrme-61,Witten-83-b,BNRS-83}. At large $N_{c}$ these models lead to
certain relations, which are not sensitive to peculiarities of different
models. These model-independent relations are believed to hold also in the
large-$N_{c}$ QCD. Many of these results can be derived using large-$N_{c}$
consistency relations \cite{GS-84,DM-93,Jenkins-93,DJM-94}.

In the approaches based on the mean-field solutions of effective theories,
one should properly treat the symmetries violated by the mean field. In
principle this problem can be solved in every order of the
$1/N_{c}$-expansion. However, sometimes the systematic large-$N_{c}$ analysis is
replaced by the rigid-rotator approximation. In the chiral limit the
rigid-rotator Hamiltonian has the simple form
\begin{equation}
H_{\mathrm{rot}}=\sum\limits_{a=1}^{3}\frac{J_{a}^{2}}{2I_{1}}
+\sum\limits_{a=4}^{7}\frac{J_{a}^{2}}{2I_{2}}\,,
\label{H-rot}
\end{equation}
where $J_{a}$ are generators of the $SU(3)$ group and $I_{k}$ are the
``moments of inertia'' of the rotator. Although the structure of the
effective rigid-rotator Hamiltonian (\ref{H-rot}) is inspired by the
large-$N_{c}$ limit, the rigid-rotator approximation is not equivalent to the
systematic $1/N_{c}$-expansion. Therefore the results based on the
rigid-rotator approximation should be taken with certain caution.

The analysis of exotic states in Refs. \cite{Praszalowicz-87,DPP-97} was
based on the effective Hamiltonian (\ref{H-rot}). The moments of inertia
$I_{k}$ grow with $N_{c}$: 
\begin{equation}
I_{k}=O(N_{c})\,.
\end{equation}
Therefore the low-lying states described by the Hamiltonian (\ref{H-rot})
have the excitation energy $O(N_{c}^{-1})$. These low-lying $O(N_{c}^{-1})$
excitations have nonexotic quantum numbers coinciding with the quantum
numbers which can be obtained in the naive quark model with $N_{c}$ colors.

In Ref. \cite{DP-03} it is argued that the rigid-rotator Hamiltonian
(\ref{H-rot}) can be applied to describe exotic states whose excitation energy is 
$O(N_{c}^{0})$. On the other hand, the authors of
Refs.~\cite{Cohen-03a,Cohen-Lebed-03,CDNL-03,IKOR-03} claim that the extension of the
rigid-rotator approach to the exotic sector is inconsistent with the
systematic $1/N_{c}$-expansion.

The argumentation of Ref. \cite{DP-03} includes a toy model of exotic states
(electrically charged quantum particle moving in the field of a static
monopole) which was studied earlier in Ref.~\cite{Guadagnini-84}.
Within this toy model
the relevance of the rigid-rotator approximation for the description of the
exotic states was explicitly checked in Ref. \cite{DP-03}. On the other
hand, the objections of Refs. \cite{Cohen-03a,Cohen-Lebed-03,CDNL-03,IKOR-03}
are mainly based on the fact that the $O(N_{c}^{0})$-excitations should be
described by the quadratic form of fluctuations around the mean field and
not by the effective rotational Hamiltonian.

In order to clarify this controversy, we have chosen several toy models
which reproduce many properties of QCD, including the large-$N_{c}$ counting
for baryons. In these models we compare the rigid-rotator approximation with
the systematic $1/N_{c}$-expansion.

One of our models is exactly solvable. In contrast to the toy model
considered in Ref. \cite{DP-03}, the exactly solvable model studied in this
paper reveals several inconsistencies of the rigid-rotator approximation. In
short, these problems can be described as follows.

1) The rigid-rotator approximation produces correct quantum numbers for the
low-lying (nonexotic) $O(N_{c}^{-1})$ excitations but leads to wrong values
of the energy.

2) The spectrum of the rigid-rotator Hamiltonian contains $O(N_{c}^{0})$
exotic excitations which are absent in the exact solution of the toy model.

One can also construct a systematic $1/N_{c}$-expansion without using the
exact solution. In this paper we compute the two first orders of the
$1/N_{c} $-expansion for the energy spectrum. We also find the spin-dependent
part of the next-to-next-to-leading (NNL) order. Since our model contains
fermion variables, the equations appearing in the systematic
$1/N_{c}$-expansion are quite similar to the well-known equations of the many-body
physics \cite{RS-80}. In particular, the leading $O(N_{c})$ order of the
$1/N_{c}$-expansion is described by the Hartree equation. The next-to-leading
(NL) $O(N_{c}^{0})$ order leads to the large-$N_{c}$ version of the RPA
(random phase approximation) equations. The spin dependent part of the NNL
$O(N_{c}^{-1})$ order is described by the Thouless--Valatin formula
\cite{TV-62}, whereas the rigid-rotator approximation leads to the Inglis formula
\cite{Inglis-54} for the same quantity, which shows the incompatibility of
the rigid-rotator approximation with the $1/N_{c}$-expansion.

The main part of the paper is devoted to the models containing colored
fermions. In addition we consider another class of models based on local
actions with color-singlet boson fields. We show that in a general case the
rigid-rotator approximation does not work in the exotic sector of these
models either.

The structure of the paper is as follows. In Section~\ref{rigid-rotator-section}, a brief description of the spectrum of the
rigid-rotator Hamiltonian is given. In Section~\ref{Hartree-RPA-section} we
describe the general structure of the $1/N_{c}$-expansion for models with
colored fermions in terms of Hartree and RPA equations.
In Section~\ref{toy-model-secton} we introduce an exactly solvable toy model with colored
fermions. The exact solution of this model is described in
Section~\ref{exact-solution-section}.
In Section~\ref{Large-N-section} we construct the
$1/N_{c}$-expansion for the energy spectrum of the toy model in the leading
and next-to-leading order. The spectrum in the NNL order is studied in
Section~\ref{rotational-excitations-section}. The discrepancy between the
rigid-rotator model and the systematic $1/N_{c}$-expansion is discussed in
Section~\ref{rigid-rotator-problems-section}.
In Section~\ref{local-theories-section} we revisit the toy model of
Refs.~\cite{DP-03,Guadagnini-84} where the rigid-rotator approximation
was successful,
we show that this success is accidental. In Conclusions we summarize the
results obtained in the toy models and make comments on the relevance of
these results for the exotic baryons in large-$N_{c}$ QCD.

\section{Rigid-rotator approximation}

\label{rigid-rotator-section}

We begin with a brief review of the properties of the rigid-rotator
Hamiltonian (\ref{H-rot}). This Hamiltonian acts in the space of the wave
functions $\psi (R)$ depending on the $SU(3)$ matrices $R$. Operators $J_{a}$
appearing in Eq.~(\ref{H-rot}) are generators of the right $SU(3)$ rotations 
\begin{equation}
\left[ \exp \left( i\omega _{a}J_{a}\right) \psi \right] (R)=\psi \left[
R\,\exp (i\omega _{a}\lambda _{a}/2)\right] \,\quad (a=1,2,\ldots ,8).
\end{equation}
Here $\lambda_a$ are Gell-Mann matrices.
The components $J_{1,2,3}$ are directly interpreted as spin operators,
whereas the spectrum of the component $J_{8}$ is constrained by the
condition \cite{Witten-83-b}
\begin{equation}
J_{8}=-\frac{N_{c}B}{2\sqrt{3}}\,,  \label{J8-constrint}
\end{equation}
where $B$ is the baryon charge.

The left rotations with generators $T_{a}$ 
\begin{equation}
\left[ \exp \left( i\omega _{a}T_{a}\right) \psi \right] (R)=\psi \left[
\exp (-i\omega _{a}\lambda _{a}/2)R\,\right]
\end{equation}
are interpreted as $SU(3)_{\mathrm{fl}}$ flavor transformations.

In the space of the rigid-rotator wave functions $\psi (R)$ one has the
identity 
\begin{equation}
\sum\limits_{a=1}^{8}T_{a}T_{a}=\sum\limits_{a=1}^{8}J_{a}J_{a}\,.
\end{equation}
In terms of the standard $(p,q)$ nomenclature of the irreducible $SU(3)$
representations, generators $T_{a}$ and $J_{a}$ belong to complex conjugate
representations $(p,q)$ and $(q,p)$ respectively. The Casimir operator is 
\begin{equation}
\sum\limits_{a=1}^{8}T_{a}T_{a}
=\sum\limits_{a=1}^{8}J_{a}J_{a}=C_{2}^{SU(3)}(p,q)=C_{2}^{SU(3)}(q,p) 
=\frac{1}{3}\left( p^{2}+q^{2}+pq\right) +p+q\,.  \label{C2-SU3}
\end{equation}

Together with the $J_{8}$ constraint (\ref{J8-constrint}) this leads to the
well-known problem of the $1/N_{c}$-expansion for the baryonic states in the
case of the $SU(3)_{\mathrm{fl}}$ flavor group: in the large-$N_{c}$ limit
all traditional (octet, decuplet,...) and exotic (e.g. antidecuplet)
representations of $SU(3)_{\mathrm{fl}}$ are lost. Instead, one has to work
with the large-$N_{c}$ $(p,q)$ representations~\cite{DuPr-88} imitating
octet, decuplet, antidecuplet etc. The condition (\ref{J8-constrint}) leads
to the following constraint on the allowed $(p,q)$ representations 
\begin{equation}
p+2q\geq N_{c}\,.
\end{equation}
In particular, the large-$N_{c}$ analogs of the main $SU(3)_{\mathrm{fl}}$
multiplets are

1) octet: $p=1,\,q=(N_{c}-1)/2$,

2) decuplet: $p=3,$ $q=(N_{c}-3)/2$,

3) antidecuplet: $p=0,$ $q=(N_{c}+3)/2$.

The eigenvalues of the effective Hamiltonian $H_{\mathrm{rot}}$
corresponding to the $SU(3)_{\mathrm{fl}}$ representation $(p,q)$ and to the
spin $J$ (for $B=1$ states) are 
\begin{equation}
E_{\mathrm{rot}}=\frac{1}{6I_{2}}\left[ p^{2}+q^{2}+pq+3(p+q)\right] +\left( 
\frac{1}{2I_{1}}-\frac{1}{2I_{2}}\right) J(J+1)-\frac{N_{c}^{2}}{24I_{2}}\,.
\end{equation}

Following Ref. \cite{DP-03}, one can define the ``exoticness'' parameter (in
order to avoid the confusion with the energy we replace notation $E$, used
in Ref. \cite{DP-03}, by $\xi $) 
\begin{equation}
\xi =\frac{1}{3}(p+2q-N_{c})
\label{xi-def}
\end{equation}
running over the nonnegative integer values $\xi =0,1,2,\ldots $

Then 
\begin{equation}
E_{\mathrm{rot}}=\frac{1}{2I_{2}}\left[ \frac{N_{c}}{2}(\xi +1)+3\frac{\xi }{
2}\left( \frac{\xi }{2}+1\right) +\frac{p}{2}\left( \frac{p}{2}+1\right) 
\right] 
+\left( \frac{1}{2I_{1}}-\frac{1}{2I_{2}}\right) J(J+1)\,.
\label{E-rot-exact}
\end{equation}
Let us consider the limit 
\begin{equation}
N_{c}\rightarrow \infty \,,\quad p,\xi =\mathrm{const}\,.
\label{p-xi-fixed-large-Nc}
\end{equation}
The spin $J$ is constrained by the condition 
\begin{equation}
\frac{|p-\xi |}{2}\leq J\leq \frac{p+\xi }{2}\,.
\end{equation}
In the case $\xi =0$ we have 
\begin{equation}
p+2q=N_{c}\,,\quad J=\frac{p}{2}\quad (\xi =0)  \label{nonexotic-qn}
\end{equation}
so that the spectrum (\ref{E-rot-exact}) has the form
\begin{equation}
E_{\mathrm{rot}}=\frac{N_{c}}{4I_{2}}+\frac{1}{2I_{1}}J(J+1)\quad (\xi =0)\,.
\label{E-rot-xi-0}
\end{equation}
In the case of arbitrary $\xi $ we find from Eq. (\ref{E-rot-exact}) in the
limit (\ref{p-xi-fixed-large-Nc})
\begin{equation}
E_{\mathrm{rot}}=\frac{N_{c}}{4I_{2}}\left( \xi +1\right) +O(N_{c}^{-1}).
\label{E-rot-xi}
\end{equation}
This expression shows that the low-lying states have $\xi =0$. According to
Eq.~(\ref{E-rot-xi-0}) the splitting between the $\xi =0$ states is
controlled by the $J(J+1)$ term and has the order $O(N_{c}^{-1})$. On the
contrary, the states with $\xi \neq 0$ have $O(N_{c}^{0})$ excitation
energies.

\section{$1/N_{c}$-expansion for fermion systems in terms of Hartree and RPA
equations}

\label{Hartree-RPA-section}

In this section we describe the general structure of the $1/N_{c}$-expansion
for fermionic systems with the Hamiltonian 
\begin{equation}
H=\frac{1}{2N_{c}}\sum
\limits_{A_{1}A_{2}A_{3}A_{4}}V_{A_{1}A_{2}|A_{3}A_{4}}\left(
\sum\limits_{c^{\prime }=1}^{N_{c}}\psi _{A_{1}c^{\prime }}^{+}\psi
_{A_{2}c^{\prime }}\right) \left( \sum\limits_{c=1}^{N_{c}}\psi
_{A_{3}c}^{+}\psi _{A_{4}c}\right) \,.  \label{H-app}
\end{equation}
Here $\psi _{Ac}$ are fermionic operators (``quarks'') with the ``color''
index $c=1,2,\ldots ,N_{c}$. Index $A$ covers other degrees of freedom. We
assume the standard fermion commutation relations 
\begin{equation}
\{\psi _{Ac},\psi _{A^{\prime }c^{\prime }}^{+}\}=\delta _{AA^{\prime
}}\delta _{cc^{\prime }}\,.
\end{equation}
The Hilbert space of states can be constructed by acting with the operators
$\psi _{Ac}^{+}$ on the ``vacuum'' $|0\rangle $ obeying the condition 
\begin{equation}
\psi _{Ac}|0\rangle =0\,.
\end{equation}
The hermiticity of the Hamiltonian implies that 
\begin{equation}
V_{A_{1}A_{2}|A_{3}A_{4}}^{\ast }=V_{A_{4}A_{3}|A_{2}A_{1}}.
\label{V-hermitean}
\end{equation}
We also assume 
\begin{equation}
V_{A_{1}A_{2}|A_{3}A_{4}}=V_{A_{3}A_{4}|A_{1}A_{2}}\,.  \label{V-symmetric}
\end{equation}
The interaction (\ref{H-app}) is $SU(N_{c})$ invariant. We have no local
gauge symmetry since the model has no space: the fermions are ``pinned'' at
one point. The factor of $1/N_{c}$ is separated in the RHS of
Eq.~(\ref{H-app}) in order to keep $V$ fixed in the limit $N_{c}\rightarrow \infty $.

One way to construct the $1/N_{c}$-expansion is to use the traditional
operator approach. Then one arrives at the large-$N_{c}$ version of the
Hartree and RPA equations. These equations are well known in the many-body
physics \cite{RS-80} as well as their connection with the $1/N$-expansion
\cite{GLM-65}. Alternatively one can derive the $1/N_{c}$-expansion
using the path integral approach which is described in
Appendix~\ref{RPA-path-int-appendix}.

At large $N_{c}$ the energy $E$ of the low-lying states can be expanded as
\begin{equation}
E=N_{c}E_{1}+E_{0}+N_{c}^{-1}E_{-1}+\ldots  \label{E-expansion}
\end{equation}

In the leading order of the $1/N_{c}$-expansion the ground state energy is
given by the Hartree approximation: 
\begin{equation}
E_{1}=\frac{1}{2}V_{AB|A^{\prime }B^{\prime }}P_{BA}^{\mathrm{(occ)}
}P_{B^{\prime }A^{\prime }}^{\mathrm{(occ)}}\,.  \label{E-Hartree-2}
\end{equation}
Here $P_{BA}^{\mathrm{(occ)}}$ is the projector onto the single-particle
occupied states $\phi _{B}^{(n)}$ 
\begin{equation}
P_{BA}^{\mathrm{(occ)}}=\sum\limits_{i:\,\mathrm{occ}}\phi _{B}^{(i)}\phi
_{A}^{(i)\ast }\,,  \label{P-occ-phi}
\end{equation}
which can be found by solving Hartree equations:
\begin{equation}
\pi _{AB}\phi _{B}^{(n)}=\varepsilon _{n}\phi _{B}^{(n)}\,,
\label{pi-phi-eps}
\end{equation}
\begin{equation}
\pi _{AB}=V_{AB|A^{\prime }B^{\prime }}\sum\limits_{i:\,\mathrm{occ}}\phi
_{A^{\prime }}^{(i)\ast }\phi _{B^{\prime }}^{(i)}=V_{AB|A^{\prime
}B^{\prime }}P_{B^{\prime }A^{\prime }}^{\mathrm{(occ)}}\,.  \label{pi-V-P}
\end{equation}

In order to find the $O(N_{c}^{0})$ correction to the energy of the ground
state and the energy of the $O(N_{c}^{0})$ excitations one has to solve the
large-$N_{c}$ version of RPA equations 
\begin{equation}
\sum\limits_{n:\mathrm{nonocc}}\, \sum\limits_{j:\mathrm{occ}} \left(
A_{minj}X_{nj}^{\nu }+B_{minj}Y_{nj}^{\nu }\right) =\Omega _{\nu
}X_{mi}^{\nu }\,,  \label{RPA-1}
\end{equation}
\begin{equation}
\sum\limits_{n:\mathrm{nonocc}}\, \sum\limits_{j:\mathrm{occ}} \left(
B_{minj}^{\ast }X_{nj}^{\nu }+A_{minj}^{\ast }Y_{nj}^{\nu }\right) =-\Omega
_{\nu }Y_{mi}^{\nu }\,.  \label{RPA-2}
\end{equation}
Here indices $i,j$ correspond to occupied Hartree states whereas indices
$m,n $ stand for nonoccupied states. We use the notation 
\begin{equation}
A_{minj}=(\varepsilon _{m}-\varepsilon _{i})\delta _{mn}\delta
_{ij}+W_{mijn}\,,  \label{A-def}
\end{equation}
\begin{equation}
B_{minj}=W_{minj}\,,  \label{B-def}
\end{equation}
\begin{equation}
W_{abcd}=\sum\limits_{ABCD}\phi _{A}^{(a)\ast }\phi _{B}^{(b)}\phi
_{C}^{(c)\ast }\phi _{D}^{(d)}V_{AB|CD}\,.  \label{W-def}
\end{equation}

Although equations (\ref{RPA-1}), (\ref{RPA-2}) are quite similar to the
standard RPA equations (see e.g. Ref. \cite{RS-80}), Eqs. (\ref{RPA-1}),
(\ref{RPA-2}) have several different features. First, the color indices do
not appear in Eqs. (\ref{RPA-1}), (\ref{RPA-2}). Second, we have dropped
those terms of the standard RPA equations whose contribution is
$1/N_{c}$-suppressed. Next, our RPA equations are formulated in terms of the
Hartree and not Hartree--Fock occupation picture (the Fock term is a
$1/N_{c}$-correction which can be treated separately). The path integral derivation
of Eqs. (\ref{RPA-1}), (\ref{RPA-2}) can be found in
Appendix~\ref{RPA-path-int-appendix}.

The RPA equations (\ref{RPA-1}), (\ref{RPA-2}) can be written in the compact
form
\begin{equation}
SZ^{\nu }=\Omega _{\nu }\Sigma Z^{\nu }\,,  \label{RPA-compact}
\end{equation}
where 
\begin{equation}
Z^{\nu }=\left( 
\begin{array}{c}
X^{\nu } \\ 
Y^{\nu }
\end{array}
\right) \,,\quad \Sigma =\left( 
\begin{array}{cc}
1 & 0 \\ 
0 & -1
\end{array}
\right) \,\,,
\end{equation}
\begin{equation}
S=\left( 
\begin{array}{cc}
A & B \\ 
B^{\ast } & A^{\ast }
\end{array}
\right) \,.  \label{S-RPA}
\end{equation}
Using Eqs. (\ref{A-def}) and (\ref{B-def}), we find 
\begin{equation}
S_{minj}=\left( 
\begin{array}{cc}
(\varepsilon _{m}-\varepsilon _{i})\delta _{mn}\delta _{ij}+W_{mijn} & 
W_{minj} \\ 
W_{minj}^{\ast } & (\varepsilon _{m}-\varepsilon _{i})\delta _{mn}\delta
_{ij}+W_{mijn}^{\ast }
\end{array}
\right) \,.  \label{S-W}
\end{equation}

Once the RPA equations (\ref{RPA-1}), (\ref{RPA-2}) are solved, the
$O(N_{c}^{0})$ part $E_{0}$ in the expansion (\ref{E-expansion}) of the
energy equals 
\begin{equation}
E_{0}=-\frac{1}{2}\sum\limits_{m:\mathrm{nonocc}}\sum\limits_{i:\mathrm{occ}
}(\varepsilon _{m}-\varepsilon _{i})+\sum\limits_{\nu }\Omega _{\nu }\left(
n_{\nu }+\frac{1}{2}\right) \,.  \label{E0-harmonic-appendix}
\end{equation}
In the last term on the RHS the summation is over the positive RPA
eigenfrequencies $\Omega _{\nu }>0$ (if any). The harmonic excitations are
parametrized by the integer numbers $n_{\nu }=0,1,2,\ldots $ The
expression (\ref{E0-harmonic-appendix}) for $E_{0}$ includes several contributions. In
addition to the pure RPA term it contains the Fock contribution and the
correction corresponding to the fact that our Hamiltonian (\ref{H-app}) is
not normally ordered.

The calculation of the $O(N_{c}^{-1})$ contribution to the
$1/N_{c}$-expansion (\ref{E-expansion}) requires more work. In this paper we consider
only the spin-dependent part of the $O(N_{c}^{-1})$ contribution for so
called rotational excitations. This part of the $O(N_{c}^{-1})$ correction
is given by the Thouless--Valatin formula \cite{TV-62} whose derivation in
the context of the $1/N_{c}$-expansion is considered in
Appendix~\ref{Thouless-Valatin-appendix}.

\section{Toy model}

\label{toy-model-secton}

We want to study a toy quantum mechanical model imitating the
$SU(N_{c})_{\mathrm{color}}\otimes SU(2)_{\mathrm{spin}}\otimes SU(3)_{\mathrm{fl}}$
symmetry of QCD. The Hamiltonian of this model is a generalization of the
Lipkin--Meshkov--Glick model \cite{LMG-65}: 
\[
H_{\mathrm{toy}}=\frac{1}{2N_{c}}\sum\limits_{f_{1},f_{2},f_{1}^{\prime
},f_{2}^{\prime }=1}^{3}\quad \sum\limits_{s_{1},s_{2},s_{1}^{\prime
},s_{2}^{\prime }=1}^{2}\left( \sum\limits_{c^{\prime }=1}^{N_{c}}\psi
_{f_{2}^{\prime }s_{2}^{\prime }c^{\prime }}^{+}\psi _{f_{1}^{\prime
}s_{1}^{\prime }c^{\prime }}\right) 
\]
\begin{equation}
\times V_{f_{2}^{\prime }s_{2}^{\prime }f_{1}^{\prime }s_{1}^{\prime
}|f_{2}s_{2}f_{1}s_{1}}\left( \sum\limits_{c=1}^{N_{c}}\psi
_{f_{2}s_{2}c}^{+}\psi _{f_{1}s_{1}c}\right) \,.  \label{H-toy}
\end{equation}
Here $\psi _{fsc}$ are fermionic operators (``quarks'') with the ``flavor''
index $f=1,2,3$, the ``spin'' index $s=1,2$ and the ``color'' index
$c=1,2,\ldots ,N_{c}$. They obey the standard anticommutation relations
\begin{equation}
\{\psi _{f_{1}s_{1}c_{1}},\psi _{f_{2}s_{2}c_{2}}^{+}\}=\delta
_{f_{1}f_{2}}\delta _{s_{1}s_{2}}\delta _{c_{1}c_{2}}\,.
\end{equation}
Introducing the compact multiindex notation
\begin{equation}
A=(f,s);\quad f=1,2,3;\quad s=1,2\,,  \label{A-fs}
\end{equation}
we see that model (\ref{H-toy}) belongs to the class of models (\ref{H-app}).

The $SU(2)_{\mathrm{spin}}\otimes SU(3)_{\mathrm{fl}}$ symmetry allows the
following structure of $V$ 
\begin{equation}
V_{f_{2}^{\prime }s_{2}^{\prime }f_{1}^{\prime }s_{1}^{\prime
}|f_{2}s_{2}f_{1}s_{1}}=\sum\limits_{m=1}^{2}\sum\limits_{n=1}^{2}b_{mn}
\Sigma _{s_{2}^{\prime }s_{1}^{\prime }|s_{2}s_{1}}^{m}\Phi _{f_{2}^{\prime
}f_{1}^{\prime }|f_{2}f_{1}}^{n},  \label{V-b-Sigma-Phi}
\end{equation}
where
\begin{equation}
\Sigma _{s_{2}^{\prime }s_{1}^{\prime }|s_{2}s_{1}}^{1}=\delta
_{s_{2}^{\prime }s_{1}^{\prime }}\delta _{s_{2}s_{1}}\,,
\end{equation}
\begin{equation}
\Sigma _{s_{2}^{\prime }s_{1}^{\prime }|s_{2}s_{1}}^{2}=\delta
_{s_{2}^{\prime }s_{1}}\delta _{s_{1}^{\prime }s_{2}}\,,
\end{equation}
\begin{equation}
\Phi _{f_{2}^{\prime }f_{1}^{\prime }|f_{2}f_{1}}^{1}=\delta _{f_{2}^{\prime
}f_{1}^{\prime }}\delta _{f_{2}f_{1}}\,,
\end{equation}
\begin{equation}
\Phi _{f_{2}^{\prime }f_{1}^{\prime }|f_{2}f_{1}}^{2}=\delta _{f_{2}^{\prime
}f_{1}}\delta _{f_{1}^{\prime }f_{2}}\,,
\end{equation}
and $b_{mn}$ are arbitrary coefficients. With this choice of $V$ we have 
\[
H_{\mathrm{toy}}=\frac{1}{2N_{c}}\left[ b_{11}\left( \psi _{f^{\prime
}s^{\prime }c^{\prime }}^{+}\psi _{f^{\prime }s^{\prime }c^{\prime }}\right)
\left( \psi _{fsc}^{+}\psi _{fsc}\right) +b_{12}\left( \psi _{f^{\prime
}s^{\prime }c^{\prime }}^{+}\psi _{fs^{\prime }c^{\prime }}\right) \left(
\psi _{fsc}^{+}\psi _{f^{\prime }sc}\right) \right. 
\]
\begin{equation}
\left. +b_{21}\left( \psi _{f^{\prime }s^{\prime }c^{\prime }}^{+}\psi
_{f^{\prime }sc^{\prime }}\right) \left( \psi _{fsc}^{+}\psi _{fs^{\prime
}c}\right) +b_{22}\left( \psi _{f^{\prime }s^{\prime }c^{\prime }}^{+}\psi
_{fsc^{\prime }}\right) \left( \psi _{fsc}^{+}\psi _{f^{\prime }s^{\prime
}c}\right) \right] \,.  \label{H-toy-b}
\end{equation}

\section{Exact solution of the toy model}

\label{exact-solution-section}

Let us diagonalize Hamiltonian (\ref{H-toy-b}). It is convenient to consider
the pair of indices (\ref{A-fs}) as an $SU(6)$ multiindex. As shown in
Appendix~\ref{Casimir-appendix}, the fermionic realization of the $SU(6)$
Casimir operator is
\begin{equation}
C_{2}^{SU(6)}=\frac{1}{2}\left( \psi _{f^{\prime }s^{\prime }c^{\prime
}}^{+}\psi _{fsc^{\prime }}\right) \left( \psi _{fsc}^{+}\psi _{f^{\prime
}s^{\prime }c}\right) -\frac{1}{12}\left( \psi _{f^{\prime }s^{\prime
}c^{\prime }}^{+}\psi _{f^{\prime }s^{\prime }c^{\prime }}\right) \left(
\psi _{fsc}^{+}\psi _{fsc}\right) \,.  \label{C2-SU6-fermionic}
\end{equation}
Let us define the ``fermion number'' 
\begin{equation}
N_{\psi }=\psi _{fsc}^{+}\psi _{fsc}\,,
\end{equation}
``spin''
\begin{equation}
J_{i}=\frac{1}{2}\psi _{fs^{\prime }c}^{+}(\sigma_{i})_{s^{\prime }s}\psi
_{fsc}\,  \label{J-i-def}
\end{equation}
and ``flavor'' generators
\begin{equation}
T_{a}=\frac{1}{2}\psi _{f^{\prime }sc}^{+}(\lambda _{a})_{f^{\prime }f}\psi
_{fsc}\,,
\end{equation}
where $\sigma_i$ and $\lambda_a$ are Pauli and Gell-Mann matrices respectively.
Using identity (\ref{t-t-sum}) from Appendix~\ref{Casimir-appendix},
we find
\begin{equation}
\sum\limits_{i=1}^{3}J_{i}^{2}=\frac{1}{2}\left( \psi _{f^{\prime }s^{\prime
}c^{\prime }}^{+}\psi _{f^{\prime }sc^{\prime }}\right) \left( \psi
_{fsc}^{+}\psi _{fs^{\prime }c}\right) -\frac{1}{4}N_{\psi }^{2}\,,
\end{equation}
\begin{equation}
\sum\limits_{a=1}^{8}T_{a}^{2}=\frac{1}{2}\left( \psi _{f^{\prime }s^{\prime
}c^{\prime }}^{+}\psi _{fs^{\prime }c^{\prime }}\right) \left( \psi
_{fsc}^{+}\psi _{f^{\prime }sc}\right) -\frac{1}{6}N_{\psi }^{2}\,.
\end{equation}
Now the Hamiltonian (\ref{H-toy-b}) can be rewritten in the form
\[
H_{\mathrm{toy}}=\frac{1}{2N_{c}}\left[ b_{11}N_{\psi }^{2}+b_{12}\left(
2\sum\limits_{a=1}^{8}T_{a}^{2}+\frac{1}{3}N_{\psi }^{2}\right) \right. 
\]
\begin{equation}
\left. +b_{21}\left( 2\sum\limits_{i=1}^{3}J_{i}^{2}+\frac{1}{2}N_{\psi
}^{2}\right) +b_{22}\left( 2C_{2}^{SU(6)}+\frac{1}{6}N_{\psi }^{2}\right) 
\right] \,.  \label{H-toy-2}
\end{equation}
This expression solves the problem of the diagonalization of the
Hamiltonian. Indeed, all operators appearing here,
\begin{equation}
C_{2}^{SU(6)}\,,\quad N_{\psi },\quad \sum\limits_{a=1}^{8}T_{a}^{2}\,,\quad
\sum\limits_{i=1}^{3}J_{i}^{2}\,,
\end{equation}
commute with each other.

We are interested in the ``baryonic'' color-singlet states containing $N_{c}$
quarks
\begin{equation}
\varepsilon _{c_{1}c_{2}\ldots c_{N_{c}}}\psi _{f_{1}s_{1}c_{1}}^{+}\psi
_{f_{2}s_{2}c_{2}}^{+}\ldots \psi
_{f_{N_{c}}s_{N_{c}}c_{N_{c}}}^{+}|0\rangle \,.  \label{Baryonic-states}
\end{equation}
For these states
\begin{equation}
N_{\psi }=N_{c}\,.  \label{N-psi-N-c}
\end{equation}
The states (\ref{Baryonic-states}) are totally symmetric in the $SU(6)$
multiindices $A_{k}=(f_{k}s_{k})$ and correspond to the $SU(6)$ irreducible
representation of dimension 
\begin{equation}
D=\frac{(N_{c}+5)!}{5!N_{c}!}
\end{equation}
with the eigenvalue of the Casimir operator
\begin{equation}
C_{2}^{SU(6)}=\frac{5}{12}N_{c}\left( N_{c}+6\right) \,.  \label{C2-SU6-B1}
\end{equation}
This irreducible $SU(6)$ representation has the following decomposition
in terms of  $SU(2)_{\mathrm{spin}}\otimes SU(3)_{\mathrm{fl}}$ irreducible
representations:
\begin{equation}
\bigoplus\limits_{k=0}^{[N_{c}/2]}\left[ (J^{(k)})_{SU(2)}\otimes
(p^{(k)},q^{(k)})_{SU(3)}\right] \,.
\end{equation}
Here $[N_c/2]$ stands for the integer part of $N_c/2$. The $SU(2)$ spins $J^{(k)}$ are given by
\begin{equation}
J^{(k)}=\frac{N_{c}}{2}-k\,\quad (k=0,1,\ldots ,[N_{c}/2])  \label{Jk}
\end{equation}
and the $p^{(k)},q^{(k)}$ parameters of the $SU(3)_{\mathrm{fl}}$
irreducible representations are
\begin{equation}
p^{(k)}=N_{c}-2k,\quad q^{(k)}=k\,.  \label{pk-qk}
\end{equation}
For the squared spin operator we obviously have
\begin{equation}
\sum\limits_{i=1}^{3}J_{i}^{2}=J^{(k)}(J^{(k)}+1)\,.  \label{JJ}
\end{equation}
Using Eqs.~(\ref{C2-SU3}), (\ref{Jk}), (\ref{pk-qk}), we obtain 
\begin{equation}
\left. \sum\limits_{a=1}^{8}T^{a}T^{a}\right| _{(p^{(k)},q^{(k)})}=\frac{1}{
12}N_{c}^{2}+\frac{1}{2}N_{c}+J^{(k)}(J^{(k)}+1)\,.  \label{TT-B1}
\end{equation}

Now we find the spectrum of energies of the toy model inserting
Eqs.~(\ref{N-psi-N-c}), (\ref{C2-SU6-B1}), (\ref{JJ}), (\ref{TT-B1}) into
Eq.~(\ref{H-toy-2}) 
\[
E_{\mathrm{toy}}^{(k)}=\frac{1}{2}\left\{ b_{11}N_{c}+b_{12}\left[ \frac{1}{2
}N_{c}+1+\frac{2}{N_{c}}J^{(k)}(J^{(k)}+1)\right] \right. 
\]
\begin{equation}
\left. +b_{21}\left[ \frac{2}{N_{c}}J^{(k)}(J^{(k)}+1)+\frac{1}{2}N_{c}
\right] +b_{22}\left( N_{c}+5\right) \right\} \,.  \label{E-toy-model}
\end{equation}
If we concentrate on the states with $J^{(k)}=O(N_{c}^{0})$, then the
large-$N_{c}$ structure of this exact result is as follows:
\begin{equation}
E_{\mathrm{toy}}^{(k)}=N_{c}E_{1}+E_{0}+\frac{1}{N_{c}}E_{-1}^{(k)}\,,
\label{E-toy-Nc-expansion}
\end{equation}
where 
\begin{equation}
E_{1}=\frac{1}{2}\left[ b_{11}+b_{22}+\frac{1}{2}\left( b_{12}+b_{21}\right) 
\right] \,,  \label{E1-res}
\end{equation}
\begin{equation}
E_{0}=\frac{1}{2}\left( b_{12}+5b_{22}\right) \,,  \label{E0-res}
\end{equation}
\begin{equation}
E_{-1}^{(k)}=\left( b_{12}+b_{21}\right) J^{(k)}(J^{(k)}+1)\,.
\label{rot-exact}
\end{equation}
The spins $J^{(k)}$ (\ref{Jk}) and the $SU(3)_{\mathrm{fl}}$ representations
$(p^{(k)},q^{(k)})$ (\ref{pk-qk}) of states, appearing in the spectrum of the
toy model, coincide with the quantum numbers (\ref{nonexotic-qn}) of the
nonexotic $\xi =0$ states in the rigid-rotator model. Moreover, the $J^{(k)}$
dependence of the $N_{c}^{-1}E_{-1}^{(k)}$ part of the
energy (\ref{rot-exact}) has the same $J(J+1)$ form as the spectrum of the
rigid rotator (\ref{E-rot-xi-0}).
However, the states with a nonzero ``exoticness''
(\ref{xi-def}) $\xi \neq 0$ are absent in the exact solution of the toy model.

\section{Large-$N_{c}$ approach to the toy model: leading and
next-to-leading orders}

\label{Large-N-section}

In the previous section we extracted the $1/N_{c}$-expansion
(\ref{E-toy-Nc-expansion}) for the spectrum of the toy model directly from the
exact solution (\ref{E-toy-model}). On the other hand, it is instructive to
construct the systematic $1/N_{c}$-expansion for the toy model directly,
without using the exact solution. The general formalism of the
$1/N_{c}$-expansion for fermion systems was described in
Section \ref{Hartree-RPA-section}. Now we want to apply this formalism to the toy model
(\ref{H-toy-b}).

We have to solve Hartree equations (\ref{pi-phi-eps}), (\ref{pi-V-P}). In
principle, these equations have many solutions describing various branches
of the spectrum of our model. We are interested in the solution
corresponding to the baryonic states (color-singlet states with $N_{\psi
}=N_{c}$). In order to find this solution, let us make use of the analogy
with the large-$N_{c}$ QCD. In QCD the large-$N_{c}$ baryons are described
by a mean-field solution (of some unknown effective theory) which has a
combined $SU(2)$ symmetry with respect to simultaneous space and flavor
rotations. In the toy model we are also interested in the solutions of
Eqs. (\ref{pi-phi-eps}), (\ref{pi-V-P}) which are invariant under combined
spin-flavor $SU(2)$ rotations. Let us introduce projectors $P^{(k)}$
($k=1,2,3$) compatible with this spin-flavor symmetry:
\begin{equation}
P_{f_{2}s_{2}f_{1}s_{1}}^{(1)}=\frac{1}{2}\left( \delta _{f_{1}f_{2}}\delta
_{s_{1}s_{2}}+\delta _{f_{1}s_{2}}\delta _{s_{1}f_{2}}\right) (1-\delta
_{f_{1}3})(1-\delta _{f_{2}3})\,,  \label{P1-def}
\end{equation}
\begin{equation}
P_{f_{2}s_{2}f_{1}s_{1}}^{(2)}=\frac{1}{2}\left( \delta _{f_{1}f_{2}}\delta
_{s_{1}s_{2}}-\delta _{f_{1}s_{2}}\delta _{s_{1}f_{2}}\right) (1-\delta
_{f_{1}3})(1-\delta _{f_{2}3})\,,  \label{P2-def}
\end{equation}
\begin{equation}
P_{f_{2}s_{2}f_{1}s_{1}}^{(3)}=\delta _{s_{1}s_{2}}\delta _{f_{1}3}\delta
_{f_{2}3}\,.  \label{P3-def}
\end{equation}
Using compact multiindex notation (\ref{A-fs}), we can write
\begin{equation}
\sum\limits_{B}P_{AB}^{(k)}P_{BC}^{(l)}=\delta ^{kl}P_{AC}^{(k)},
\label{P-ortho}
\end{equation}
\begin{equation}
\sum\limits_{k=1}^{3}P_{AB}^{(k)}=\delta _{AB}\,.
\end{equation}
The dimensions of these projectors can be read from their traces: 
\begin{equation}
\mathrm{Sp}P^{(1)}=3\,,\quad \mathrm{Sp}P^{(2)}=1\,,\quad \mathrm{Sp}
P^{(3)}=2\,.
\label{Sp-Pk}
\end{equation}
Now let us consider the linear combination of these projectors
\begin{equation}
\pi _{AB}=\sum\limits_{k=1}^{3}\varepsilon _{k}P_{AB}^{(k)}\,.
\label{M-P-decomposition}
\end{equation}
Combining this with Eq. (\ref{P-ortho}), we see that $\pi _{AB}$
obeys the equation
\begin{equation}
\sum_B \pi _{AB}P_{BC}^{(k)}=\varepsilon _{k}P_{AC}^{(k)}\,.
\end{equation}
Comparing this equation with Eq. (\ref{pi-phi-eps}), we conclude that
representation (\ref{M-P-decomposition}) is compatible with the Hartree
equation (\ref{pi-phi-eps}).

We are interested in the baryonic states which correspond to occupying one
level with $N_{c}$ quarks. The one-dimensional projector $P^{(2)}$ is a
perfect candidate for this occupied level. Therefore we take in
Eq. (\ref{P-occ-phi})
\begin{equation}
P_{AB}^{\mathrm{(occ)}}=\sum\limits_{i:\,\mathrm{occ}}\phi _{A}^{(i)}\phi
_{B}^{(i)\ast }=P_{AB}^{(2)}\,.  \label{Phi-Phi-occ-P2}
\end{equation}
Since the projector $P_{AB}^{(2)}$ is one-dimensional, the sum in
Eq.~(\ref{Phi-Phi-occ-P2})
contains only one occupied level: 
\begin{equation}
\phi _{fs}^{(i)}=\frac{1}{\sqrt{2}}\,\epsilon _{fs}\,\quad (\epsilon
_{12}=-\epsilon _{21}=1).
\end{equation}
Now equation (\ref{pi-V-P}) leads us to the following result:
\begin{equation}
\pi _{AB}=V_{AB|A^{\prime }B^{\prime }}P_{B^{\prime }A^{\prime }}^{(2)}\,.
\label{Pi-V-P2}
\end{equation}
The leading order of the $1/N_{c}$-expansion for the energy $E$
(\ref{E-expansion}) is given by the Hartree term (\ref{E-Hartree-2}). Using
Eq. (\ref{Phi-Phi-occ-P2}), we find
\begin{equation}
E=\frac{N_{c}}{2}P_{B^{\prime }A^{\prime }}^{(2)}V_{A^{\prime }B^{\prime
}|AB}P_{BA}^{(2)}+O(N_{c}^{0})\,.  \label{E-Hartree}
\end{equation}
With the explicit expressions for $V_{A^{\prime }B^{\prime }|AB}$
(\ref{V-b-Sigma-Phi}) and for $P^{(2)}$ (\ref{P2-def}) we easily find
\begin{equation}
E=\frac{N_{c}}{2}\left[ b_{11}+b_{22}+\frac{1}{2}\left( b_{12}+b_{21}\right) 
\right] +O(N_{c}^{0})\,.
\end{equation}
This coincides with the leading $O(N_{c})$ part (\ref{E1-res}) of the exact
solution (\ref{E-toy-Nc-expansion}).

According to Eqs. (\ref{P-ortho}), (\ref{M-P-decomposition}) the
single-particle energies are given by the following expression:
\begin{equation}
\varepsilon _{k}=\frac{\mathrm{Sp}(\pi P^{(k)})}{\mathrm{Sp}P^{(k)}}=\frac{
P_{BA}^{(2)}V_{AB|A^{\prime }B^{\prime }}P_{B^{\prime }A^{\prime }}^{(k)}}{
\mathrm{Sp}P^{(k)}}\,.
\end{equation}
Using the explicit expressions for $P^{(k)}$ (\ref{P1-def}) -- (\ref{P3-def})
and for $V$ (\ref{V-b-Sigma-Phi}), we find
\begin{equation}
\varepsilon _{1}=b_{11}+\frac{1}{2}\left( b_{12}+b_{21}\right) \,,
\label{eps-1}
\end{equation}
\begin{equation}
\varepsilon _{2}=b_{11}+b_{22}+\frac{1}{2}\left( b_{12}+b_{21}\right) \,,
\label{eps-2}
\end{equation}
\begin{equation}
\varepsilon _{3}=b_{11}+\frac{1}{2}b_{21}\,.  \label{eps-3}
\end{equation}

In the toy model (\ref{H-toy-b}) all RPA eigenfrequencies $\Omega _{\nu }$
(\ref{RPA-compact}) vanish (see Appendix \ref{RPA-toy-appendix}) so that the
harmonic excitations are absent. According to Eq. (\ref{E0-harmonic-appendix}),
in the absence of the harmonic excitations the $O(N_{c}^{0})$ correction
$E_{0}$ to the energy of the ground state is given by the equation
\begin{equation}
E_{0}=-\frac{1}{2}\sum\limits_{m:\mathrm{nonocc}}\sum\limits_{i:\mathrm{occ}
}(\varepsilon _{m}-\varepsilon _{i})\,,  \label{E0-general}
\end{equation}
where the summation runs over the single-particle occupied states
$\varepsilon _{i}$ and nonoccupied states $\varepsilon _{m}$ with the weights
corresponding to their degeneracy. In our case
according to Eq. (\ref{Sp-Pk}) the degeneracy of
non-occupied levels $\varepsilon _{1}$ and $\varepsilon _{3}$ is 3 and 2
respectively, whereas the occupied level $\varepsilon _{2}$ is not
degenerate. Therefore we find from Eqs. (\ref{eps-1}), (\ref{eps-2}),
(\ref{eps-3}), (\ref{E0-general})
\begin{equation}
E_{0}=-\frac{1}{2}\left[ 3(\varepsilon _{1}-\varepsilon _{2})+2(\varepsilon
_{3}-\varepsilon _{2})\right] =\frac{5}{2}b_{22}+\frac{1}{2}b_{12}\,.
\end{equation}
This result agrees with the expression for $E_{0}$ (\ref{E0-res}) obtained
from the exact solution.

\section{Rotational excitations in the $1/N_{c}$-expansions}

\label{rotational-excitations-section}

As mentioned above, the mean-field solution $\pi _{AB}$ (\ref{Pi-V-P2})
violates both $SU(2)_{\mathrm{spin}}$ and $SU(3)_{\mathrm{fl}}$ symmetries
but it is invariant under combined spin-flavor $SU(2)$ rotations. Therefore
the systematic $1/N_{c}$-expansion should involve a correct treatment of the
collective coordinates corresponding to the rotation of the mean field.
Since the spin rotation of the mean field is equivalent to the flavor
rotations, it is sufficient to consider only the $SU(3)$ flavor rotations.
The invariance of the mean field under the $\lambda _{8}$ transformations
leads to the $J_{8}$ constraint~(\ref{J8-constrint}).

The dependence of the energy on the spin $J$ appears in the order
$O(N_{c}^{-1})$ and has the form
\begin{equation}
\Delta E_{J}=\frac{J(J+1)}{2I}\,,  \label{J-I-term}
\end{equation}
where $I=O(N_{c})$. In the framework of the systematic $1/N_{c}$-expansion
the ``moment of inertia'' $I$ can be computed using the large-$N_{c}$
version of the Thouless--Valatin formula \cite{TV-62}
(see Appendix~\ref{Thouless-Valatin-appendix}). Applying this
formula to the toy model, we find
[see Eqs. (\ref{I-coorect-app}),
(\ref{S-a-equation}) in Appendix \ref{I-calc-appendix}]
\begin{equation}
I=\frac{N_{c}}{4}\sum\limits_{i_{1}i_{2}:\mathrm{occ}}\,\sum
\limits_{m_{1}m_{2}:\mathrm{nonocc}}\left( 
\begin{array}{c}
\langle i_{1}|\lambda _{3}|m_{1}\rangle \\ 
\langle m_{1}|\lambda _{3}|i_{1}\rangle
\end{array}
\right) ^{T}\left( S^{-1}\right) _{m_{1}i_{1}m_{2}i_{2}}\left( 
\begin{array}{c}
\langle m_{2}|\lambda _{3}|i_{2}\rangle \\ 
\langle i_{2}|\lambda _{3}|m_{2}\rangle
\end{array}
\right) \,.  \label{I-correct}
\end{equation}
Here $S$ is the matrix (\ref{S-RPA}) entering the large-$N_{c}$ RPA equation
(\ref{RPA-compact}). Strictly speaking, matrix $S$ is degenerate so that
expression (\ref{I-correct}) has to be rewritten in a more careful form. The
details can be found in Appendix \ref{I-calc-appendix}. The resulting
expression for $I$ in the toy model (\ref{H-toy-b}) is given by
Eq. (\ref{I-correct-res})
\begin{equation}
I=\frac{N_{c}}{2\left( b_{12}+b_{21}\right) }\,.  \label{I-correct-res-again}
\end{equation}
Inserting this expression into Eq. (\ref{J-I-term}), we obtain
\begin{equation}
\Delta E_{J}=\frac{b_{12}+b_{21}}{N_{c}}J(J+1)\,,
\end{equation}
which agrees with the $N_{c}^{-1}$ part (\ref{rot-exact}) of the exact
expression for the energy.

\section{Problems of the rigid-rotator approximation}

\label{rigid-rotator-problems-section}

The expression (\ref{I-correct}) for the moment of inertia $I$
requires the knowledge of the RPA matrix $S$. The detailed description of
this matrix is given in Section \ref{Hartree-RPA-section}
[see Eq. (\ref{S-W})]. If one omits the direct $V$ contribution to $S$ [represented by the $W$
in the RHS of Eq. (\ref{S-W})] then one arrives at the following
approximation for $S$:
\begin{equation}
S_{m_{1}i_{1}m_{2}i_{2}}\approx \delta _{m_{1}m_{2}}\delta
_{i_{1}i_{2}}(\varepsilon _{m_{1}}-\varepsilon _{i_{1}})\,.
\label{S-e-approx}
\end{equation}
We stress that this approximation is not justified by the large-$N_{c}$
limit.

Inserting this approximate expression into Eq. (\ref{I-correct}), one obtains
the Inglis expression for the moment of inertia \cite{Inglis-54}
\begin{equation}
I_{1}=\frac{N_{c}}{2}\sum\limits_{i:\mathrm{occ}}\,\sum\limits_{m:\mathrm{
nonocc}}\frac{\langle i|\lambda _{3}|m\rangle \langle m|\lambda
_{3}|i\rangle }{\varepsilon _{m}-\varepsilon _{i}}\,.
\label{I1-Inglis}
\end{equation}

This expression can be also ``derived'' in the rigid-rotator approximation
if one concentrates only on the rotational degrees of freedom instead of the
systematic $1/N_{c}$-expansion (see Appendix \ref{Rigid-rotator-appendix}
and Ref. \cite{DPP-88}). In this way one arrives at the matrix
(\ref{I-ab-app}) 
\begin{equation}
I_{ab}=\frac{N_{c}}{4}\sum\limits_{i:\,\mathrm{occ}}\,\,\sum\limits_{m:\,
\mathrm{nonocc}}\frac{\langle i|\lambda _{a}|m\rangle \langle m|\lambda
_{b}|i\rangle +\langle i|\lambda _{b}|m\rangle \langle m|\lambda
_{a}|i\rangle }{\varepsilon _{m}-\varepsilon _{i}}\,.  \label{I-general-2}
\end{equation}
Here the occupied state $i$ is associated with the projector $P^{(2)}$
(\ref{P2-def}), whereas the nonoccupied states $m$ correspond to the projectors
$P^{(1)}$ (\ref{P1-def}) and $P^{(3)}$ (\ref{P3-def}). Due to the $SU(2)$
spin-flavor symmetry of the mean field, the matrix $I_{ab}$ is diagonal: 
\begin{equation}
I_{1}\equiv I_{11}=I_{22}=I_{33}\,,
\end{equation}
\begin{equation}
I_{2}\equiv I_{44}=I_{55}=I_{66}=I_{77}\,.
\end{equation}

The transition $\varepsilon _{2}\leftrightarrow \varepsilon _{1}$ in
Eq. (\ref{I-general-2}) contributes to the moment of inertia $I_{1}$, whereas the
transition $\varepsilon _{2}\leftrightarrow \varepsilon _{3}$ generates
$I_{2}$. We find from Eq. (\ref{I-general-2}): 
\begin{equation}
I_{ab}=\frac{N_{c}}{4}\sum\limits_{m=1,3}\left. 
\frac{\mathrm{Sp} (P^{(i)}\lambda _{a}P^{(m)}\lambda _{b} 
+P^{(i)}\lambda_{b}P^{(m)}\lambda _{a})}
{\varepsilon _{m}-\varepsilon _{i}}
\,\right| _{i=2}\,,
\end{equation}
which results in 
\begin{equation}
I_{1}=\frac{N_{c}}{2(-b_{22})}\,,  \label{I1-res}
\end{equation}
\begin{equation}
I_{2}=\frac{N_{c}}{2(-2b_{22}-b_{12})}\,\,.
\end{equation}
In the rigid-rotator approximation the spectrum of low-lying $O(N_{c}^{-1})$
excitations is described by Eq. (\ref{E-rot-xi-0}):
\begin{equation}
\frac{J(J+1)}{2I_{1}}\,.
\end{equation}
This expression has the same structure as Eq. (\ref{J-I-term}) derived in
the systematic $1/N_{c}$-expansion. However, comparing the rigid-rotator
moment of inertia $I_{1}$ (\ref{I1-res}) with the correct expression $I$
(\ref{I-correct-res-again}), we see that the rigid-rotator approximation
gives a wrong result:
\begin{equation}
I_{1}\neq I\,.
\end{equation}
The reason is obvious: approximation (\ref{S-e-approx}) is not justified by
the large-$N_{c}$ limit. Approximation (\ref{S-e-approx})
has the same problem in the case of the model with the flavor group $SU(2)$,
see Appendix~\ref{SU2-model-appendix}.
Since the Inglis formula
(\ref{I1-Inglis}) fails already in the
nonexotic sector of $O(N_{c}^{-1})$ excitations, there is no sense to
discuss the results based on the Inglis expression for $I_{2}$. Moreover, as
mentioned above, the states of the rigid-rotator Hamiltonian~(\ref{H-rot})
with a nonzero ``exoticness''~(\ref{xi-def}) $\xi \neq 0$ are simply absent
in the spectrum of the exact solution of the toy model (\ref{H-toy-b}).

\section{Large-$N_{c}$ models with color-singlet boson fields}

\label{local-theories-section}

The large-$N_{c}$ counting in QCD is compatible with the large-$N_{c}$
behavior in various models: naive quark model, Skyrme model
\cite{Skyrme-61,Witten-83-b}, Nambu--Jona-Lasinio model (NJL) \cite{NJL-61},
chiral quark-soliton model (CQSM) \cite{DPP-88} etc. Some of these models
(e.g. NJL) contain only colored fermion fields, others (e.g. Skyrme model)
involve only color-singlet bosonic fields, some models such as CQSM contain
both color-singlet boson fields and colored quark fields. Let us concentrate
on two extreme cases of models with local (in time) actions containing

1) colored fermion fields,

2) color-singlet boson fields.

In spite of the difference between these two groups, the analysis of the
$1/N_{c}$-expansion has many common features. Indeed, we can bosonize the
interaction between the fermions and integrate out the fermions
(see Appendix~\ref{RPA-path-int-appendix}).
In this way
one arrives at a theory with color-singlet boson field $\pi $ but the action 
$S[\pi ]$ of this theory is \emph{nonlocal.}

Whatever action $S[\pi ]$ we have, local or nonlocal, the spectrum of
$O(N_{c}^{0})$ excitations
\begin{equation}
E={\rm const}+\sum_\nu \Omega_\nu n_\nu+O(N_c^{-1})\,, \qquad\qquad n_\nu=0,1,2,\ldots
\end{equation}
is given by frequencies $\omega =\Omega _{\nu }$,
which are determined by the equation
\begin{equation}
\det K(\omega )=0\,.  \label{det-K-0-0}
\end{equation}
Here $K(\omega )$ is the quadratic form of the boson action $S[\pi ]$ in the
frequency representation:
\begin{equation}
K(\omega )=\frac{1}{N_{c}}\int dte^{-i\omega t}\frac{\delta ^{2}S[\pi ]}{
\delta \pi (t)\delta \pi (0)}.
\end{equation}
In theories with nonlocal actions $S[\pi ]$, obtained via a bosonization of
local theories with colored fermions, one can reduce equation
(\ref{det-K-0-0}) to RPA equations as shown in Appendix
\ref{RPA-path-int-appendix}.

In theories with local actions $S[\pi ]$ quadratic in time derivatives, we
have the following structure of the quadratic form $K(\omega )$:
\begin{equation}
K(\omega )=-K_{2}\omega ^{2}+iK_{1}\omega +K_{0}\,.
\end{equation}
In this case Eq. (\ref{det-K-0-0}) is equivalent to the existence of a
periodic solution of the equation
\begin{equation}
\left( K_{2}\partial _{t}^{2}+K_{1}\partial _{t}+K_{0}\right) \delta \pi
(t)=0  \label{K-delta-pi}
\end{equation}
with the frequency $\omega $. This equation has a simple physical meaning:
it describes periodic solutions of the classical equations of motion
corresponding to an infinitesimal deviation from the static solution. In a
general case this problem cannot be reduced to the diagonalization of the
rigid-rotator Hamiltonian.

However, in Ref. \cite{DP-03} (see also Ref.~\cite{Guadagnini-84}) a simple model
was considered whose spectrum of $O(N_{c}^{0})$ excitations contains a
branch which can be successfully described by the rigid-rotator
approximation. This model involves a particle with mass $\mu $ and electric
charge $e$, moving in the 3-dimensional space in the field $\mathbf{H}$ of a
monopole with magnetic charge $g$
\begin{equation}
\mathbf{H}=g\frac{\mathbf{x}}{|\mathbf{x}|^{3}}=\nabla \times \mathbf{A}+[
\mathrm{string\,field}]\,.
\end{equation}
In addition the model has a central potential $V$ so that the total
Lagrangian is
\begin{equation}
L=\frac{\mu \mathbf{\dot{x}}^{2}}{2}-e(\mathbf{\dot{x}A})-V(|\mathbf{x}|)\,.
\label{L-monopole}
\end{equation}
If one interprets this Lagrangian as a sort of effective theory for
large-$N_{c}$ ``baryons'' coming from some ``microscopic'' theory, then the
large-$N_{c}$ limit corresponds to the semiclassical limit for the Lagrangian
(\ref{L-monopole}):
\begin{equation}
\mu =O(N_{c}),\quad V=O(N_{c})\,,  \label{mu-Nc}
\end{equation}
\begin{equation}
eg=\frac{1}{2}N_{c}B\,,\quad B=O(N_{c}^{0})\,.  \label{eg-B}
\end{equation}
Here the integer number $B$ has the meaning of the ``baryon charge'' in the
underlying ``microscopic'' large-$N_{c}$ theory. The representation
(\ref{eg-B}) for $eg$ agrees with the Dirac quantization rule for the magnetic
charge.

As shown in Ref. \cite{DP-03}, this large-$N_{c}$ inspired semiclassical
limit justifies the reduction of the model to the rigid-rotator Lagrangian
\begin{equation}
L_{\mathrm{rot}}=\int dt\left[ \frac{I}{2}\left( \Omega _{1}^{2}+\Omega
_{2}^{2}\right) +eg\Omega _{3}\right] \,.  \label{L-rot-monopole}
\end{equation}
for a subset of $O(N_{c}^{0})$ excitations. Here $\Omega _{a}$ is the
angular velocity. The moment of inertia $I$ is given by the equation
\begin{equation}
I=\mu R^{2}\,,\quad V^{\prime }(R)=0.  \label{I-mu-r}
\end{equation}

The Lagrangian (\ref{L-rot-monopole}) corresponds to the Hamiltonian
\begin{equation}
H_{\mathrm{rot}}=\frac{1}{2I}(J_{1}^{2}+J_{2}^{2})  \label{H-rot-monopole}
\end{equation}
with the quantization constraint
\begin{equation}
J_{3}^{\prime }=eg=\frac{N_{c}B}{2}\,.  \label{J3-Witten}
\end{equation}
Here $J_{3}^{\prime }$ is the projection of the momentum $J$ on the third
axis in the ``body-fixed'' frame. The spectrum of $J$ is
\begin{equation}
J=\frac{N_{c}|B|}{2},\frac{N_{c}|B|}{2}+1,\frac{N_{c}|B|}{2}+2,\ldots
\label{rigid-rotator-spectrum-FM}
\end{equation}
As a result the spectrum of energies is
\begin{equation}
E=\frac{1}{2I}\left[ J(J+1)-\left( J_{3}^{\prime }\right) ^{2}\right] =\frac{
1}{2I}\left[ J(J+1)-\left( \frac{N_{c}B}{2}\right) ^{2}\right]\,.
\label{H-J-3-prime}
\end{equation}
By analogy with Eq. (\ref{xi-def}) one can introduce the ``exoticness''
\begin{equation}
\xi =J-\frac{N_{c}|B|}{2}=0,1,2,3,\ldots
\end{equation}
Then one finds from Eq. (\ref{H-J-3-prime})
\begin{equation}
E=E_{\xi }=\frac{1}{2I}\left[ \frac{N_{c}|B|}{2}(2\xi +1)+\xi (\xi +1)\right]
\,.
\end{equation}
In the limit $I\sim N_{c}\rightarrow \infty $ we have equidistant
$O(N_{c}^{0})$ excitations parametrized by $\xi $:
\begin{equation}
E_{\xi }=\frac{N_{c}|B|}{4I}(2\xi +1)+O(1/N_{c})\,.
\label{E-exotic-monopole}
\end{equation}
As shown in Ref. \cite{DP-03}, the ``exotic'' spectrum
(\ref{E-exotic-monopole}) of the rigid rotator (\ref{L-rot-monopole}) coincides
(up to $1/N_{c}$ corrections) with a part of the true spectrum of the
original model (\ref{L-monopole}). This observation was used as an argument
in favor of the validity of the rigid-rotator approach to exotic states.

Let us check that the spectrum (\ref{E-exotic-monopole}) is described by one
of the frequencies $\omega $ determined by the general
equation~(\ref{det-K-0-0}). The full spectrum of the harmonic excitations can be found
from the solutions of equation (\ref{K-delta-pi}). Due to the symmetry of
the problem, the equation of motion (\ref{K-delta-pi}) can be split into two
parts. In the case of radial infinitesimal oscillations $\delta x_{rad}(t)$
controlled by the central potential, one can ignore the $K_{1}$ term
associated with the magnetic field:
\begin{equation}
\left( K_{2}\partial _{t}^{2}+K_{0}\right) \delta x_{rad}(t)=0\,.
\label{x-rad-eq}
\end{equation}
We also have the periodic motion $\delta x_{circ}(t)$ along an infinitesimal
circle due to the magnetic force. In this case we can omit the term $K_{0}$
containing the central potential:
\begin{equation}
\left( K_{2}\partial _{t}^{2}+K_{1}\partial _{t}\right) \delta
x_{circ}(t)=0\,.  \label{circ-eq}
\end{equation}
In the frequency representation Eq. (\ref{circ-eq}) takes the form
\begin{equation}
\left( -\omega ^{2}K_{2}+i\omega K_{1}\right) \delta \tilde{x}_{circ}=0\,.
\end{equation}
This equation has nonzero solutions $\delta \tilde{x}_{circ}$ for two
frequencies $\omega $. The zero frequency $\omega =0$ corresponds to the
$J_{3}$ degeneracy of the quantum spectrum. The second nonzero frequency
\begin{equation}
\omega =\frac{eH}{\mu }
\end{equation}
describes the circular motion of the particle with mass $\mu $ and charge $e$
in the magnetic field $H$. This frequency is responsible for the ``exotic
excitations'' (\ref{E-exotic-monopole}). Indeed, the magnetic field of the
monopole is $H=g/R^{2}$, and the moment of inertia $I$ is given by
Eq. (\ref{I-mu-r}) so that
\begin{equation}
\omega =\frac{eg}{I}\,.
\end{equation}
Using Eqs. (\ref{mu-Nc}), (\ref{eg-B}), (\ref{I-mu-r}), we find
\begin{equation}
\omega =\frac{N_{c}|B|}{2I}=O(N_{c}^{0})\,.
\end{equation}
This frequency agrees with the spectrum of harmonic
excitations (\ref{E-exotic-monopole}):
\begin{equation}
E_{\xi +1}-E_{\xi }=\frac{N_{c}|B|}{2I}+O(1/N_{c})\,.
\end{equation}
In addition one has excitations determined by the eigenfrequency of
Eq. (\ref{x-rad-eq}).

Thus in the case of the simple model (\ref{L-monopole}) the rigid-rotator
approximation leads to the spectrum which is a part of the full
$O(N_{c}^{0}) $ spectrum determined by Eq. (\ref{det-K-0-0}). However, this
agreement of the rigid-rotator spectrum with the general equation
(\ref{det-K-0-0}) is ``accidental''. In a general case one has to solve equation
(\ref{det-K-0-0}) or the equivalent problem (\ref{K-delta-pi}) in order to
find the spectrum of $O(N_{c}^{0})$ excitations and this solution has
nothing to do with the rigid rotator approximation. In
Appendix \ref{Extended-Guadagnini-model} we describe an extension of the model
(\ref{L-monopole}) which explicitly demonstrates the breakdown of the
rigid-rotator approximation.

Let us summarize. Both in the models with colored fermions and in the models
with color-singlet bosons the spectrum of the $O(N_{c}^{0})$ excitations is
determined by the solutions of equation (\ref{det-K-0-0}). In a general
case the rigid-rotator approximation cannot describe these excitations.

\section{Conclusions}

In this paper we have considered several simple quantum-mechanical models
imitating the large-$N_{c}$ QCD. The main aim was to understand the status
of exotic baryons at large $N_{c}$. Our results confirm the conservative
opinion that the exotic states are only a part of the $O(N_{c}^{0})$
excitations whose properties are determined by the quadratic form of the
action written in terms of color-singlet boson fields. Recent attempts to
extract reliable information about exotic baryons directly from the
rigid-rotator Hamiltonian are incompatible with the results of our analysis.

We have considered two different classes of models. The first group includes
models containing colored fermions. The second class of models is based on
local actions for color-singlet fields. In both cases the spectrum of the
$O(N_{c}^{0})$ excitations is determined by the eigenfrequencies of the
quadratic form of the action. But in the case of colored fermions one has to
work with the nonlocal bosonized action. Our analysis of this nonlocal
quadratic form was based on the reduction of the problem to RPA equations.
In the case of models with local actions for color-singlet bosons it is
convenient to perform the analysis of the quadratic form in terms of the
semiclassical quantization of periodic classical solutions.

We have studied an exactly solvable toy model with colored fermions.
We have shown that in the large-$N_{c}$ limit this model is described by a mean-field
solution whose spin-flavor symmetry coincides with the symmetry of the
large-$N_{c}$ baryons in QCD. Using this mean-field solution as a starting point
for the systematic $1/N_{c}$-expansion, we have computed the energy spectrum
in the first three orders of the $1/N_{c}$-expansion. The results of the
$1/N_{c}$-expansion are in complete agreement with the exact spectrum of the
model.

We have also tested the rigid-rotator approximation. The
results coming from the rigid-rotator approximation are rather
disappointing. The effective rotational Hamiltonian derived in the context
of the rigid-rotator approximation has exotic baryonic states which are
absent in the exact solution of the model.

The rigid-rotator approximation also has problems in the nonexotic sector.
Although the rigid-rotator approximation predicts the correct quantum
numbers for the nonexotic low-lying $O(N_{c}^{-1})$ baryonic excitations,
the values of the energy coming from the rigid-rotator approximation are
wrong. The comparison of the rigid-rotator approximation with the systematic 
$1/N_{c}$-expansion allows us to detect the origin of the problem: the
rigid-rotator approximation leads to the Inglis formula which ignores the
quadratic form of fluctuations around the mean field. On the contrary, the
calculation of the rotational spectrum based on the systematic
$1/N_{c}$-expansion results in the Thouless--Valatin formula which involves the
coefficients of the RPA equations.

Although our model with colored fermions reproduces the main symmetries of
QCD, at one point there is an important difference: the absence of the
exotic states in the toy model is simply a model artifact: the ``baryonic''
states in the model contain only $N_{c}$ quarks above the bare vacuum. This
leaves no possibility to have extra quark-antiquark pairs which could
generate exotic quantum numbers. In QCD the situation is different: the
existence of exotic states is allowed. However, the question is whether the
rigid-rotator approximation can \emph{correctly} describe the exotic baryon
states in QCD. Our analysis of the toy model raises doubts about the
relevance of the rigid-rotator approximation. Actually it is not a problem
to construct solvable models containing exotic states and to show explicitly
that the rigid-rotator approximation does not describe properly these
exotic states.

We also have considered a group of models with local actions for
color-singlet boson fields. This group contains a simple model compatible
with the rigid-rotator approximation. However, our analysis shows that this
agreement is a consequence of the simplicity of the model. A slight
extension of this model leads to the breakdown of the rigid-rotator
approximation in the exotic sector.

To summarize, our results agree with the traditional approach to
the $O(N_{c}^{0})$ baryonic excitations (including the exotic ones)
as states controlled by the quadratic form of fluctuations of the effective
action. Within the framework of the large-$N_{c}$ expansion in QCD (without
extra model assumptions) we see no
justification of the rigid-rotator approximation.

Certainly, one should be rather careful with the extension of our toy-model
experience to the case of the large-$N_{c}$ QCD. For example, the fermionic
toy model is based on the quantum mechanics of spin systems in 0-dimensional
space. This model has colored fermions but no gluons and no gauge symmetry.
In the model calculations we were interested in the discrete eigenstates
of the Hamiltonian, whereas in reality exotic baryons are resonances.
One also should keep in mind that this toy model imitates the strict chiral
limit of QCD. In the real world the situation is more difficult since one
has to deal with a nontrivial interplay between two small parameters: the
light-quark mass matrix and $1/N_{c}$.

\textbf{Acknowledgments.} The author appreciates many interesting and
instructive discussions with Ya. I. Azimov, T.~D.~Cohen, D.~I.~Diakonov,
V.~Yu.~Pet\-rov and M.~V.~Polyakov. This work was supported by DFG and BMBF.

\appendix\renewcommand{\theequation}{\Alph{section}.\arabic{equation}}

\section{Path integral derivation of Hartree and RPA equations in the
large-$N_{c}$ expansion}

\label{RPA-path-int-appendix}

In this appendix we use the path integral approach in order to construct the 
$1/N_{c}$-expansion for models with four-fermionic interaction and show how
Hartree and RPA equations appear in the context of the $1/N_{c}$-expansion.

We start from the Lagrangian corresponding to the model Hamiltonian
(\ref{H-app})
\begin{equation}
L=i\psi _{Ac}^{+}\partial _{t}\psi _{Ac}-\frac{1}{2N_{c}}\left( \psi
_{A^{\prime }c^{\prime }}^{+}\psi _{B^{\prime }c^{\prime }}\right)
V_{A^{\prime }B^{\prime }|AB}\left( \psi _{Ac}^{+}\psi _{Bc}\right) \,.
\end{equation}
One can bosonize the four-fermion interaction, introducing auxiliary
variables $\pi _{AB}$ and integrating out fermion variables $\psi ,\psi ^{+}$
:
\[
\int D\psi \int D\psi ^{+}\exp \left\{ i\int dt\left[ i\psi
_{Ac}^{+}\partial _{t}\psi _{Ac}-\frac{1}{2N_{c}}\left( \psi _{A^{\prime
}c^{\prime }}^{+}\psi _{B^{\prime }c^{\prime }}\right) V_{A^{\prime
}B^{\prime }|AB}\left( \psi _{Ac}^{+}\psi _{Bc}\right) \right] \right\} 
\]
\[
=\int D\pi \int D\psi \int D\psi ^{+}\exp \left\{ i\int dt\left[ i\psi
_{Ac}^{+}\partial _{t}\psi _{Ac}-\psi _{Ac}^{+}\pi _{AB}\psi _{Bc}+\frac{
N_{c}}{2}\pi _{A^{\prime }B^{\prime }}(V^{-1})_{A^{\prime }B^{\prime
}|AB}\pi _{AB}\right] \right\} 
\]
\begin{equation}
=\int D\pi \exp iS[\pi ]\,.
\end{equation}
Here we have the effective action
\begin{equation}
S[\pi ]=N_{c}\left[ -i\ln \det (i\partial _{t}-\pi )+\frac{1}{2}\int dt\,\pi
_{A^{\prime }B^{\prime }}(V^{-1})_{A^{\prime }B^{\prime }|AB}\pi _{AB}\right]
\,.  \label{S-bosonized}
\end{equation}

The presence of the factor $N_{c}$ in this action allows us to apply the
saddle point method for the construction of the $1/N_{c}$-expansion. The
saddle point equation is
\begin{equation}
\frac{1}{N_{c}}\frac{\delta S[\pi ]}{\delta \pi _{AB}(t)}=i\langle t,B|\frac{
1}{i\partial _{t}-\pi }|t,A\rangle +(V^{-1})_{AB|A^{\prime }B^{\prime }}\pi
_{A^{\prime }B^{\prime }}=0\,.  \label{delta-S-saddle-point}
\end{equation}
For static saddle points $\pi $ we have
\begin{equation}
\langle t,B|\frac{1}{i\partial _{t}-\pi }|t,A\rangle =\int \frac{d\omega }{
2\pi }\sum\limits_{n}\frac{\phi _{B}^{(n)}\phi _{A}^{(n)\ast }}{\omega
-\varepsilon _{n}}  \label{t-diag}
\end{equation}
where $\phi _{B}^{(n)}$ and $\varepsilon _{n}$ are eigenvectors and
eigenvalues of the matrix $\pi :$
\begin{equation}
\pi _{AB}\phi _{B}^{(n)}=\varepsilon _{n}\phi _{B}^{(n)}\,.
\label{pi-epsilon}
\end{equation}
The $\pm i0$ pole uncertainties in Eq. (\ref{t-diag}) should be resolved
relying on the physical picture of occupied states
\begin{equation}
-i\int \frac{d\omega }{2\pi }\sum\limits_{n}\frac{\phi _{B}^{(n)}\phi
_{A}^{(n)\ast }}{\omega -\varepsilon _{n}}=\sum\limits_{i:\mathrm{occ}}\phi
_{B}^{(i)}\phi _{A}^{(i)\ast }\equiv P_{BA}^{\mathrm{(occ)}}.
\label{int-omega-occ}
\end{equation}
After the $\omega $ integration, the summation runs only over the occupied
single-particle states $\phi ^{(i)}$.

Inserting Eqs. (\ref{t-diag}) and (\ref{int-omega-occ}) into the
saddle-point equation (\ref{delta-S-saddle-point}), we obtain the Hartree
equation
\begin{equation}
\pi _{AB}=V_{AB|A^{\prime }B^{\prime }}P_{B^{\prime }A^{\prime }}^{\mathrm{
(occ)}}\,.  \label{pi-Hartree}
\end{equation}
Computing the action (\ref{S-bosonized}) for this $\pi _{AB}$, one easily
finds the $O(N_{c})$ Hartree energy (\ref{E-Hartree-2})
\begin{equation}
E=\frac{N_{c}}{2}V_{AB|A^{\prime }B^{\prime }}P_{BA}^{\mathrm{(occ)}
}P_{B^{\prime }A^{\prime }}^{\mathrm{(occ)}}\,+O(N_{c}^{0})\,.
\end{equation}

Now we want to compute the $O(N_{c}^{0})$ corrections. In a general case
these corrections will describe the harmonic excitations with some
frequencies $\Omega _{\nu }$
\begin{equation}
\Delta E=\sum\limits_{\nu }n_{\nu }\Omega _{\nu }=O(N_{c}^{0})\,,
\label{Delta-E-c-Omega}
\end{equation}
where numbers $n_{\nu }=0,1,2,\ldots $ are integer. In order to find these
frequencies we must consider the quadratic form of the action
(\ref{S-bosonized}) in the frequency representation:
\begin{equation}
K_{A_{1}B_{1}|A_{2}B_{2}}(\omega )=\int dt_{1}e^{-i\omega t_{1}}\frac{\delta
^{2}}{\delta \pi _{A_{1}B_{1}}(t_{1})\delta \pi _{A_{2}B_{2}}(0)}\frac{1}{
N_{c}}S[\pi ]\,.  \label{K-omega-def}
\end{equation}
The spectrum of the harmonic excitations (\ref{Delta-E-c-Omega}) is
determined by the frequencies $\omega =\Omega _{\nu }$ at which the
quadratic form $K$ becomes degenerate:
\begin{equation}
\det K(\Omega _{\nu })=0\,.  \label{det-K-0}
\end{equation}

The calculation of the quadratic form of the action (\ref{S-bosonized}) is
straightforward:
\[
\frac{1}{N_{c}}\frac{\delta ^{2}}{\delta \pi _{A_{1}B_{1}}(t_{1})\delta \pi
_{A_{2}B_{2}}(t_{2})}S[\pi ] 
\]
\begin{equation}
=i\langle t_{2},B_{2}|\frac{1}{i\partial _{t}-\pi }|t_{1},A_{1}\rangle
\langle t_{1},B_{1}|\frac{1}{i\partial _{t}-\pi }|t_{2},A_{2}\rangle +\left(
V^{-1}\right) _{A_{1}B_{1}|A_{2}B_{2}}\delta (t_{1}-t_{2})\,.
\end{equation}
Now the quadratic form (\ref{K-omega-def}) can be represented as follows:
\begin{equation}
K_{A_{1}B_{1}|A_{2}B_{2}}(\omega )=iL_{A_{1}B_{1}|A_{2}B_{2}}(\omega
)+\left( V^{-1}\right) _{A_{1}B_{1}|A_{2}B_{2}}\,,  \label{K-L-V}
\end{equation}
where
\[
L_{A_{1}B_{1}|A_{2}B_{2}}(\omega )\equiv \int dt_{1}e^{-i\omega t_{1}}\int
\langle 0,B_{2}|\frac{1}{i\partial _{t}-\pi }|t_{1},A_{1}\rangle \langle
t_{1},B_{1}|\frac{1}{i\partial _{t}-\pi }|0,A_{2}\rangle 
\]
\begin{equation}
=-i\sum\limits_{i:\mathrm{occ}}\,\sum\limits_{m:\mathrm{nonocc}}\left[ \frac{
\langle B_{2}|i\rangle \langle i|A_{1}\rangle \langle B_{1}|m\rangle \langle
m|A_{2}\rangle }{\varepsilon _{m}-\varepsilon _{i}-\omega -i0}+\frac{\langle
B_{2}|m\rangle \langle m|A_{1}\rangle \langle B_{1}|i\rangle \langle
i|A_{2}\rangle }{\varepsilon _{m}-\varepsilon _{i}+\omega -i0}\right] \,.
\label{L-def}
\end{equation}
Here we use the standard Dirac notation
\begin{equation}
\langle A | k \rangle
=\langle k | A \rangle^\ast
=\phi_A^{(k)}\,.
\end{equation}
Taking some antiunitary operator $\theta $ (e.g. time reversal)
\begin{equation}
|A^{\theta }\rangle =\theta |A\rangle ,\quad \langle A^{\theta }|B^{\theta
}\rangle =\langle B|A\rangle \,,
\end{equation}
\begin{equation}
\pi^{\theta }=\theta \pi\theta ^{-1}\,
\end{equation}
and introducing compact notation
\begin{equation}
|AB\rangle =|A\rangle \otimes |B\rangle \,,
\end{equation}
\begin{equation}
D=\pi\otimes 1-1\otimes \pi^{\theta }\,,  \label{D-H-def}
\end{equation}
we can write
\begin{equation}
\langle i|A_{1}\rangle \langle B_{1}|m\rangle =\langle B_{1}|m\rangle
\langle A_{1}^{\theta }|i^{\theta }\rangle =\langle B_{1}A_{1}^{\theta
}|mi^{\theta }\rangle \,,
\end{equation}
\begin{equation}
D|mi^{\theta }\rangle =(\varepsilon _{m}-\varepsilon _{i})|mi^{\theta
}\rangle \,.  \label{D-mi-theta}
\end{equation}
Then expression (\ref{L-def}) takes the form
\begin{equation}
L_{A_{1}B_{1}|A_{2}B_{2}}(\omega )=-i\langle B_{1}A_{1}^{\theta }|\left(
P_{1}\frac{1}{D-\omega -i0}-P_{2}\frac{1}{D-\omega +i0}\right)
|A_{2}B_{2}^{\theta }\rangle \,.  \label{L-calc-1}
\end{equation}
Here we have introduced projectors
\begin{equation}
P_{1}=\sum\limits_{i:\mathrm{occ}}\,\sum\limits_{m:\mathrm{nonocc}
}|mi^{\theta }\rangle \langle mi^{\theta }|\,,  \label{P1-RPA-def}
\end{equation}
\begin{equation}
P_{2}=\sum\limits_{i:\mathrm{occ}}\,\sum\limits_{m:\mathrm{nonocc}
}|im^{\theta }\rangle \langle im^{\theta }|\,,  \label{P2-RPA-def}
\end{equation}
\begin{equation}
P_{k}P_{l}=\delta _{kl}P_{l}\,.
\end{equation}
Next we define
\begin{equation}
\Sigma =P_{1}-P_{2}\,,  \label{Sigma-P-def}
\end{equation}
\begin{equation}
Q(\omega )=\Sigma \frac{1}{D-\omega -i0\cdot \Sigma }\,.  \label{Q-def}
\end{equation}
Then the RHS of Eq. (\ref{L-calc-1}) takes the simple form
\begin{equation}
L_{A_{1}B_{1}|A_{2}B_{2}}(\omega )=-i\langle B_{1}A_{1}^{\theta }|Q(\omega
)|A_{2}B_{2}^{\theta }\rangle \,.
\end{equation}
Inserting this into Eq. (\ref{K-L-V}), we see that
\begin{equation}
K_{A_{1}B_{1}|A_{2}B_{2}}(\omega )=\langle B_{1}A_{1}^{\theta }|Q(\omega
)|A_{2}B_{2}^{\theta }\rangle +\left( V^{-1}\right)
_{A_{1}B_{1}|A_{2}B_{2}}\,.  \label{K-Q}
\end{equation}
Although $\theta $ is an antiunitary operator, for any given basis
$\{|A\rangle \}$ we can find such a unitary operator $U$ that (for the basis
states $|A\rangle $ only)
\begin{equation}
|A^{\theta }\rangle =U|A\rangle \,.  \label{theta-U}
\end{equation}
Now expression (\ref{K-Q}) becomes
\begin{equation}
K_{A_{1}B_{1}|A_{2}B_{2}}(\omega )=\langle B_{1}A_{1}|(1\otimes
U^{+})Q(\omega )(1\otimes U)|A_{2}B_{2}\rangle +\left( V^{-1}\right)
_{A_{1}B_{1}|A_{2}B_{2}}\,.  \label{K-U-Q}
\end{equation}
Next we define the permutation operator $\Pi $
\begin{equation}
\Pi |AB\rangle =|BA\rangle \,.
\end{equation}
This allows us to rewrite relation (\ref{K-U-Q}) in the operator form
without indices $A_{k},B_{k}$
\begin{equation}
K(\omega )=\Pi (1\otimes U^{+})Q(\omega )(1\otimes U)+V^{-1}\,.
\label{K-U-Q-2}
\end{equation}
Let us define
\begin{equation}
\tilde{K}(\omega )=(1\otimes U)\Pi K(\omega )(1\otimes U^{+})\,,
\label{K-tilde-def}
\end{equation}
\begin{equation}
\tilde{V}=(1\otimes U)V\Pi (1\otimes U^{+})\,.  \label{V-tilde-def}
\end{equation}
Now it follows from Eq. (\ref{K-U-Q-2})
\begin{equation}
\tilde{K}(\omega )=Q(\omega )+\tilde{V}^{-1}\,.  \label{K-tilde-Q-tilde}
\end{equation}
The next step is to define the projector
\begin{equation}
P=P_{1}+P_{2}\,,  \label{P-P12-def}
\end{equation}
\begin{equation}
P^{2}=P
\end{equation}
and to notice that 
\begin{equation}
Q=PQ=QP=PQP\,.  \label{PQ-id}
\end{equation}
The operator $Q$ can be inverted on the subspace of the projector $P$. Let
us consider operator $P\left( Q^{-1}+\tilde{V}\right) P$ on this subspace
and let $R$ be its inverse:
\begin{equation}
R\left[ P\left( Q^{-1}+\tilde{V}\right) P\right] =\left[ P\left( Q^{-1}+
\tilde{V}\right) P\right] R=P\,,  \label{R-inv-def}
\end{equation}
\begin{equation}
RP=PR=R\,.  \label{R-PR}
\end{equation}
For brevity we will write
\begin{equation}
R=\left[ P\left( Q^{-1}+\tilde{V}\right) P\right] ^{-1}=P\left( Q^{-1}+P
\tilde{V}P\right) ^{-1}P\,,  \label{R-short}
\end{equation}
implying that all ambiguities should be understood in the sense of
Eqs. (\ref{R-inv-def}), (\ref{R-PR}).

Multiplying equality (\ref{R-inv-def}) by $Q$ and using Eqs. (\ref{PQ-id}),
(\ref{R-PR}), we find
\begin{equation}
R+Q\tilde{V}R=Q\,.  \label{R-Q-V-id}
\end{equation}
Using Eq. (\ref{K-tilde-Q-tilde}), we can write
\begin{equation}
\tilde{K}\left( \tilde{V}-\tilde{V}R\tilde{V}\right) =\left( Q+\tilde{V}
^{-1}\right) \left( \tilde{V}-\tilde{V}R\tilde{V}\right) =1-R\tilde{V}
+\left( Q-Q\tilde{V}R\right) \tilde{V}\,.
\end{equation}
We can simplify the RHS using Eq. (\ref{R-Q-V-id}):
\begin{equation}
\tilde{K}\left( \tilde{V}-\tilde{V}R\tilde{V}\right) =1\,.
\end{equation}
Combining this with Eq. (\ref{R-short}), we find
\begin{equation}
\tilde{K}^{-1}=\tilde{V}-\tilde{V}R\tilde{V}=\tilde{V}-\tilde{V}P\left(
Q^{-1}+P\tilde{V}P\right) ^{-1}P\tilde{V}\,.
\end{equation}
Inserting Eq. (\ref{Q-def}), we arrive at
\begin{equation}
\tilde{K}^{-1}(\omega )=\tilde{V}-\tilde{V}P\left[ \Sigma D+P\tilde{V}
P-\omega \Sigma -i0\right] ^{-1}P\tilde{V}\,.  \label{K-tilde-inv}
\end{equation}

According to Eq. (\ref{K-tilde-def}) we have
\begin{equation}
|\det K(\omega )|=|\det \tilde{K}(\omega )|\,.
\end{equation}
Therefore Eq. (\ref{det-K-0}) is equivalent to the equation
\begin{equation}
\det \tilde{K}(\Omega _{\nu })=0\,.
\end{equation}
Combining this with Eq. (\ref{K-tilde-inv}), we see that the problem reduces
to the equation
\begin{equation}
\det \left( \Sigma D+P\tilde{V}P-\Omega _{\nu }\Sigma \right) =0\,,
\end{equation}
where the determinant must be computed on the subspace of the projector $P$.
This is equivalent to the eigenvalue equation
\begin{equation}
SZ^{\nu }=\Omega _{\nu }\Sigma Z^{\nu }\,,  \label{RPA-derived}
\end{equation}
where
\begin{equation}
S=\Sigma D+P\tilde{V}P\,.
\label{S-Sigma-PVP}
\end{equation}

Equation (\ref{RPA-derived}) is nothing else but RPA equation
(\ref{RPA-compact}). Indeed, according to Eqs. (\ref{D-H-def}),
(\ref{D-mi-theta}), (\ref{P1-RPA-def}), (\ref{P2-RPA-def}), (\ref{Sigma-P-def}) we have
\[
\Sigma D=\sum\limits_{i:\mathrm{occ}}\,\sum\limits_{m:\mathrm{nonocc}}\left(
|mi^{\theta }\rangle \langle mi^{\theta }|-|im^{\theta }\rangle \langle
im^{\theta }|\right) \left( \pi\otimes 1-1\otimes \pi^{\theta }\right) 
\]
\begin{equation}
=\sum\limits_{i:\mathrm{occ}}\,\sum\limits_{m:\mathrm{nonocc}}(\varepsilon
_{m}-\varepsilon _{i})\left( |mi^{\theta }\rangle \langle mi^{\theta
}|+|im^{\theta }\rangle \langle im^{\theta }|\right) \,.
\label{Sigma-D-explicit}
\end{equation}
Using Eqs. (\ref{P1-RPA-def}), (\ref{P2-RPA-def}), (\ref{theta-U}),
(\ref{V-tilde-def}), (\ref{P-P12-def}), we can write
\[
P\tilde{V}P=\sum\limits_{i,j:\mathrm{occ}}\,\sum\limits_{m,n:\mathrm{nonocc}
}\left( |mi^{\theta }\rangle \langle mi^{\theta }|+|im^{\theta }\rangle
\langle im^{\theta }|\right) 
\]
\[
\times (1\otimes U)V\Pi (1\otimes U^{+})\left( |nj^{\theta }\rangle \langle
nj^{\theta }|+|jn^{\theta }\rangle \langle jn^{\theta }|\right) 
\]
\begin{equation}
=\sum\limits_{i,j:\mathrm{occ}}\,\sum\limits_{m,n:\mathrm{nonocc}}\left(
|mi^{\theta }\rangle \langle mi|+|im^{\theta }\rangle \langle im|\right)
V\left( |jn\rangle \langle nj^{\theta }|+|nj\rangle \langle jn^{\theta
}|\right) \,.
\end{equation}
Now we use Eq. (\ref{W-def})
\[
P\tilde{V}P=\sum\limits_{i,j:\mathrm{occ}}\sum\limits_{m,n:\mathrm{nonocc}}
\left[ |mi^{\theta }\rangle W_{mijn}\langle nj^{\theta }|+|im^{\theta
}\rangle W_{imjn}\langle nj^{\theta }|\right. 
\]
\begin{equation}
\left. +|mi^{\theta }\rangle W_{minj}\langle jn^{\theta }|+|im^{\theta
}\rangle W_{imnj}\langle jn^{\theta }|\right] \,.  \label{PVP-explicit}
\end{equation}
Taking the vector $Z^{\nu }$ in the form
\begin{equation}
Z^{\nu }=\sum\limits_{i:\mathrm{occ}}\sum\limits_{m:\mathrm{nonocc}}\left(
X_{mi}|mi^{\theta }\rangle +Y_{mi}|im^{\theta }\rangle \right) \,,
\end{equation}
we find using Eqs. (\ref{Sigma-P-def}), (\ref{Sigma-D-explicit}),
(\ref{PVP-explicit})
\begin{equation}
\Sigma Z^{\nu }=\sum\limits_{i:\mathrm{occ}}\sum\limits_{m:\mathrm{nonocc}
}\left( X_{mi}^{\nu }|mi^{\theta }\rangle -Y_{mi}^{\nu }|im^{\theta }\rangle
\right) \,,
\end{equation}
\begin{equation}
(\Sigma D)Z^{\nu }=\sum\limits_{i:\mathrm{occ}}\sum\limits_{m:\mathrm{nonocc}
}(\varepsilon _{m}-\varepsilon _{i})\left( X_{mi}^{\nu }|mi^{\theta }\rangle
+Y_{mi}^{\nu }|im^{\theta }\rangle \right) \,,
\end{equation}
\[
\left( P\tilde{V}P\right) Z^{\nu }=\sum\limits_{i,j:\mathrm{occ}
}\sum\limits_{m,n:\mathrm{nonocc}}\left[ |mi^{\theta }\rangle \left(
W_{mijn}X_{nj}^{\nu }+W_{minj}Y_{nj}^{\nu }\right) \right. 
\]
\begin{equation}
\left. +|im^{\theta }\rangle \left( W_{imjn}X_{nj}^{\nu
}+W_{imnj}Y_{nj}^{\nu }\right) \right] \,.
\end{equation}
Inserting these expressions into Eqs. (\ref{RPA-derived}), (\ref{S-Sigma-PVP})
we obtain
\begin{equation}
(\varepsilon _{m}-\varepsilon _{i})X_{mi}^{\nu }+\sum\limits_{j:\mathrm{occ}
}\sum\limits_{n:\mathrm{nonocc}}\left( W_{mijn}X_{nj}^{\nu
}+W_{minj}Y_{nj}^{\nu }\right) =\Omega _{\nu }X_{mi}^{\nu }\,,
\end{equation}
\begin{equation}
(\varepsilon _{m}-\varepsilon _{i})Y_{mi}^{\nu }+\sum\limits_{j:\mathrm{occ}
}\sum\limits_{n:\mathrm{nonocc}}\left( W_{imjn}X_{nj}^{\nu
}+W_{imnj}Y_{nj}^{\nu }\right) =-\Omega _{\nu }Y_{mi}^{\nu }\,.
\end{equation}
Using relations (\ref{V-hermitean}), (\ref{V-symmetric}), we see that these
equations coincide with RPA equations (\ref{RPA-1}), (\ref{RPA-2}).

\section{Thouless--Valatin formula at large $N_{c}$}

\label{Thouless-Valatin-appendix}

In this appendix we derive the large-$N_{c}$ version of the Thouless--Valatin
formula \cite{TV-62} for the moment of inertia controlling $O(N_{c}^{-1})$
rotational excitations (\ref{J-I-term}). This formula is well known in the
many-body physics \cite{RS-80}. The derivation presented below follows the
traditional ideas but emphasizes the role of the large $N_{c}$ in the
justification of Thouless--Valatin formula. We start from a rotationally
invariant Hamiltonian $H$ commuting with the angular momentum~$J_{k}$ 
\begin{equation}
\lbrack H,J_{k}]=0\,.  \label{H-J-commute}
\end{equation}
In the case of the rotational excitations (\ref{J-I-term}) the dependence on
the spin $J$ appears in the order $O(N_{c}^{-1})$. In principle, the
calculation of the NNLO $O(N_{c}^{-1})$ corrections to the energy requires a
serious work. But if one is interested only in the $J$ dependent part of
this $O(N_{c}^{-1})$ contribution, then the result can be computed using
Thouless--Valatin formula.

Let us consider the auxiliary Hamiltonian $H-\omega J_{3}$ instead of the
original Hamiltonian $H$ and let us study the ground state energy
$E^{(0)}(\omega )$ of the Hamiltonian $H-\omega J_{3}$ as a function of $
\omega $ in the interval
\begin{equation}
N_{c}^{-1}\ll \omega \ll N_{c}^{0}\,.  \label{omega-interval}
\end{equation}
In this intermediate region we can treat $\omega $ either as a ``numerically
small'' $O(N_{c}^{0})$ quantity or as a ``numerically large'' $O(N_{c}^{-1})$
quantity. Both approaches allow us to derive corresponding expressions for
the ground state energy $E^{(0)}(\omega )$. Matching the results in the
intermediate region (\ref{omega-interval}), one can obtain a simple
representation for the moment of inertia $I$ controlling the $J(J+1)$
splitting (\ref{J-I-term}).

If we consider the case of $\omega =O(N_{c}^{0})$, then the ground state
energy $E^{(0)}(\omega )$ of the Hamiltonian $H-\omega J_{3}$ is given by
the $\omega $ dependent Hartree energy $N_{c}E_{1}(\omega )$
\begin{equation}
E^{(0)}(\omega )=N_{c}E_{1}(\omega )+O(N_{c}^{0})\,.
\end{equation}
Let us compute the second order derivative of this expression with respect
to $\omega $ in the interval (\ref{omega-interval}):
\begin{equation}
\left. \frac{\partial ^{2}}{\partial \omega ^{2}}E^{(0)}(\omega )\right|
_{N_{c}^{-1}\ll \omega \ll 1}\approx \frac{\partial ^{2}}{\partial \omega
^{2}}N_{c}E_{1}(\omega )\approx \left. \frac{\partial ^{2}}{\partial \omega
^{\prime 2}}N_{c}E_{1}(\omega ^{\prime })\right| _{\omega ^{\prime }=0}\,.
\label{dE-derivative}
\end{equation}

Now let us change the approach and consider $\omega $ as a ``numerically
large'' $O(N_{c}^{-1})$ quantity. We start from the $1/N_{c}$-expansion for
the Hamiltonian $H$ (without the $\omega J_{3}$ correction)
\begin{equation}
E_{J}^{(0)}=N_{c}E_{1}+E_{0}+\frac{1}{N_{c}}\left[ c_{1}J(J+1)+c_{2}\right]
+O(N_{c}^{-2})\,.  \label{E-0-J}
\end{equation}
According to Eq. (\ref{H-J-commute}) Hamiltonian $H$ commutes with $J_{3}$.
Therefore we find from Eq. (\ref{E-0-J}) for the energy $E^{(0)}(\omega )$
of the ground state of the Hamiltonian $H-\omega J_{3}$
\[
E^{(0)}(\omega )=\min_{J,J_{3}}\left( E_{J}^{(0)}-\omega J_{3}\right) 
\]
\begin{equation}
=\min_{J,J_{3}}\left\{ N_{c}E_{1}+E_{0}+\frac{1}{N_{c}}\left[
c_{1}J(J+1)+c_{2}\right] -\omega J_{3}+O(N_{c}^{-2})\right\} \,.
\end{equation}
Obviously the minimum is achieved at $J_{3}=J$ so that
\begin{equation}
E^{(0)}(\omega )=\min_{J}\left\{ N_{c}E_{1}+E_{0}+\frac{1}{N_{c}}\left[
c_{1}J(J+1)+c_{2}\right] -\omega J+O(N_{c}^{-2})\right\} \,.
\label{E-omega-min-J}
\end{equation}
The point of the minimum is
\begin{equation}
J=\frac{\omega N_{c}-c_{1}}{2c_{1}}\,.
\end{equation}
Due to condition (\ref{omega-interval}) we have
\begin{equation}
1\ll J\approx \frac{\omega N_{c}}{2c_{1}}\ll N_{c}\,.
\end{equation}
Now we find from Eq. (\ref{E-omega-min-J})
\begin{equation}
E^{(0)}(\omega )\approx N_{c}E_{1}+E_{0}-\frac{\omega ^{2}N_{c}}{4c_{1}}\,.
\end{equation}
We see that
\begin{equation}
\left. \frac{\partial ^{2}}{\partial \omega ^{2}}E^{(0)}(\omega )\right|
_{N_{c}^{-1}\ll \omega \ll 1}\approx -\frac{N_{c}}{2c_{1}}\,.
\end{equation}
Combining this result with Eq. (\ref{dE-derivative}), we see that
\begin{equation}
\left. \frac{\partial ^{2}}{\partial \omega ^{\prime 2}}N_{c}E_{1}(\omega
^{\prime })\right| _{\omega ^{\prime }=0}=-\frac{N_{c}}{2c_{1}}\,.
\label{d2E-c1}
\end{equation}
Comparing Eqs. (\ref{E-0-J}) and (\ref{J-I-term}) we find the connection
between the coefficient $c_{1}$ and the moment of inertia $I$:
\begin{equation}
c_{1}=\frac{N_{c}}{2I}\,.
\end{equation}
Now we find from Eq. (\ref{d2E-c1})
\begin{equation}
I=-N_{c}\left. \frac{\partial ^{2}}{\partial \omega ^{2}}E_{1}\left( \omega
\right) \right| _{\omega =0}\,.  \label{I-E-Hartree}
\end{equation}

This intermediate result simplifies the calculation of the moment of inertia 
$I$. Indeed, in the beginning the problem of the calculation of $I$ was a
NNLO problem. But now it is reduced to the expression (\ref{I-E-Hartree})
containing only the Hartree energy $N_{c}E_{1}\left( \omega \right) $ for
the Hamiltonian $H-\omega J_{3}$, i.e. to the problem of the leading order
$O(N_{c})$. The calculation of the second order derivative with respect to
$\omega $ in Eq. (\ref{I-E-Hartree}) is equivalent to the construction of the
second perturbation theory in $\omega $ for the Hartree equation.

Now we have to write the Hartree equation for the Hamiltonian $H-\omega
J_{3} $ with
\begin{equation}
J_{3}=\frac{1}{2}\psi ^{+}\sigma _{3}\psi \,,
\end{equation}
where $\sigma _{3}$ is the Pauli matrix acting on the spin indices. Due to
the $\omega J_{3}$ term the expression for the Hartree
energy~(\ref{E-Hartree-2}) is modified as follows
\begin{equation}
E_{1}=\frac{1}{2}V_{AB|A^{\prime }B^{\prime }}P_{BA}^{\mathrm{(occ)}
}P_{B^{\prime }A^{\prime }}^{\mathrm{(occ)}}-\frac{\omega }{2}\left( \sigma
_{3}\right) _{AB}P_{BA}^{(occ)}.  \label{E-Hartree-3}
\end{equation}
Let us differentiate this expression with respect to $\omega $:
\begin{equation}
\frac{\partial E_{1}}{\partial \omega }=h_{AB}\frac{\partial P_{BA}^{\mathrm{
(occ)}}}{\partial \omega }-\frac{1}{2}\left( \sigma _{3}\right) _{AB}P_{BA}^{
\mathrm{(occ)}}\,.  \label{dE1-calc-1}
\end{equation}
Here
\begin{equation}
h_{AB}=V_{AB|A^{\prime }B^{\prime }}P_{B^{\prime }A^{\prime }}^{\mathrm{(occ)
}}-\frac{\omega }{2}\left( \sigma _{3}\right) _{AB}  \label{h-V-omega}
\end{equation}
is nothing else but the one-particle Hartree Hamiltonian with the single
particle eigenfunctions
\begin{equation}
h_{AB}\phi _{B}^{(n)}=\varepsilon _{n}\phi _{B}^{(n)}  \label{h-phi-eps-3}
\end{equation}
and with the projector onto the occupied states
\begin{equation}
P_{BA}^{\mathrm{(occ)}}=\sum\limits_{i:\,\mathrm{occ}}\phi _{B}^{(i)}\phi
_{A}^{(i)\ast }\,.
\end{equation}
The projector $P_{BA}^{\mathrm{(occ)}}$ depends on the parameter $\omega $
of the Hamiltonian $H-\omega J_{3}$. The standard perturbation theory allows
us to compute the derivative
\[
\frac{\partial P_{BA}^{\mathrm{(occ)}}}{\partial \omega }=\sum\limits_{i:
\mathrm{occ}}\,\sum\limits_{m:\mathrm{nonocc}}\frac{1}{\varepsilon
_{i}-\varepsilon _{m}} 
\]
\qquad \qquad 
\begin{equation}
\times \left[ \langle m|\frac{\partial h}{\partial \omega }|i\rangle \phi
_{B}^{(m)}\phi _{A}^{(i)\ast }+\langle i|\frac{\partial h}{\partial \omega }
|m\rangle \phi _{B}^{(i)}\phi _{A}^{(m)\ast }\right] \,.  \label{dP-d-omega}
\end{equation}
Combining this with Eq. (\ref{h-phi-eps-3}), we see that
\begin{equation}
h_{AB}\frac{\partial P_{BA}^{\mathrm{(occ)}}}{\partial \omega }=0\,.
\end{equation}
We insert this result into Eq. (\ref{dE1-calc-1})
\begin{equation}
\frac{\partial E_{1}}{\partial \omega }=-\frac{1}{2}\left( \sigma
_{3}\right) _{AB}P_{BA}^{\mathrm{(occ)}}.
\end{equation}
Next we take the second $\omega $ derivative
\begin{equation}
\frac{\partial ^{2}E_{1}}{\partial \omega ^{2}}=-\frac{1}{2}\left( \sigma
_{3}\right) _{AB}\frac{\partial P_{BA}^{\mathrm{(occ)}}}{\partial \omega }.
\end{equation}
Using Eq. (\ref{dP-d-omega}), we find
\[
\frac{\partial ^{2}E_{1}}{\partial \omega ^{2}}=-\frac{1}{2}\sum\limits_{i:
\mathrm{occ}}\,\sum\limits_{m:\mathrm{nonocc}}\frac{1}{\varepsilon
_{i}-\varepsilon _{m}} 
\]
\qquad \qquad 
\begin{equation}
\times \left[ \langle m|\frac{\partial h}{\partial \omega }|i\rangle \langle
i|\sigma _{3}|m\rangle +\langle i|\frac{\partial h}{\partial \omega }
|m\rangle \langle m|\sigma _{3}|i\rangle \right] \,.
\label{d2-E-d-omega-calc-1}
\end{equation}
According to Eq. (\ref{h-V-omega}) we have
\begin{equation}
\frac{\partial h_{AB}}{\partial \omega }=V_{AB|A^{\prime }B^{\prime }}\frac{
\partial P_{B^{\prime }A^{\prime }}^{\mathrm{(occ)}}}{\partial \omega }-
\frac{1}{2}\left( \sigma _{3}\right) _{AB}\,.
\end{equation}
Inserting expression (\ref{dP-d-omega}), we find
\[
\frac{\partial h_{AB}}{\partial \omega }=-\frac{1}{2}\left( \sigma
_{3}\right) _{AB}+V_{AB|A^{\prime }B^{\prime }}\sum\limits_{j:\mathrm{occ}
}\,\sum\limits_{n:\mathrm{nonocc}}\frac{1}{\varepsilon _{j}-\varepsilon _{n}}
\]
\qquad \qquad 
\begin{equation}
\times \left[ \langle n|\frac{\partial h}{\partial \omega }|j\rangle \phi
_{B^{\prime }}^{(n)}\phi _{A^{\prime }}^{(j)\ast }+\langle j|\frac{\partial h
}{\partial \omega }|n\rangle \phi _{B^{\prime }}^{(j)}\phi _{A^{\prime
}}^{(n)\ast }\right] \,.
\end{equation}
Now we take the matrix element of this equation
\[
\langle m|\frac{\partial h}{\partial \omega }|i\rangle =-\frac{1}{2}\langle
m|\sigma _{3}|i\rangle 
\]
\begin{equation}
+\sum\limits_{j:\mathrm{occ}}\,\sum\limits_{n:\mathrm{nonocc}}\frac{1}{
\varepsilon _{j}-\varepsilon _{n}}\left[ W_{mijn}\langle n|\frac{\partial h}{
\partial \omega }|j\rangle +W_{minj}\langle j|\frac{\partial h}{\partial
\omega }|n\rangle \right] \,.  \label{m-dh-i}
\end{equation}
In the case when the state $i$ is occupied and $m$ is nonoccupied, let us
introduce the notation
\begin{equation}
\tilde{a}_{mi}=-\frac{2}{\varepsilon _{m}-\varepsilon _{i}}\langle m|\frac{
\partial h}{\partial \omega }|i\rangle \,,
\end{equation}
\begin{equation}
\tilde{b}_{mi}=-\frac{2}{\varepsilon _{m}-\varepsilon _{i}}\langle i|\frac{
\partial h}{\partial \omega }|m\rangle \,.
\end{equation}
Then we find from Eq. (\ref{m-dh-i})
\begin{equation}
\left( \varepsilon _{m}-\varepsilon _{i}\right) \tilde{a}_{mi}=\langle
m|\sigma _{3}|i\rangle -\sum\limits_{j:\mathrm{occ}}\,\sum\limits_{n:\mathrm{
nonocc}}\left( W_{mijn}\tilde{a}_{nj}+W_{minj}\tilde{b}_{nj}\right) \,.
\end{equation}
Exchanging $i\leftrightarrow m$ in Eq. (\ref{m-dh-i}), we obtain the second
equation
\begin{equation}
\left( \varepsilon _{m}-\varepsilon _{i}\right) \tilde{b}_{mi}=\langle
i|\sigma _{3}|m\rangle -\sum\limits_{j:\mathrm{occ}}\,\sum\limits_{n:\mathrm{
nonocc}}\left( W_{imjn}\tilde{a}_{nj}+W_{imnj}\tilde{b}_{nj}\right) \,.
\end{equation}
Obviously these equations can be rewritten in the form
\begin{equation}
\sum\limits_{i^{\prime }:\mathrm{occ}}\,\sum\limits_{m^{\prime }:\mathrm{
nonocc}}S_{mim^{\prime }i^{\prime }}\left( 
\begin{array}{c}
\tilde{a}_{m^{\prime }i^{\prime }} \\ 
\tilde{b}_{m^{\prime }i^{\prime }}
\end{array}
\right) =\left( 
\begin{array}{c}
\langle m|\sigma _{3}|i\rangle \\ 
\langle i|\sigma _{3}|m\rangle
\end{array}
\right) \,,  \label{S-a-equation-0}
\end{equation}
where $S_{mim^{\prime }i^{\prime }}$ is the RPA matrix (\ref{S-W}).

Now we find from Eq. (\ref{d2-E-d-omega-calc-1})
\begin{equation}
\frac{\partial ^{2}E_{1}}{\partial \omega ^{2}}=-\frac{1}{4}\sum\limits_{i:
\mathrm{occ}}\,\sum\limits_{m:\mathrm{nonocc}}\left( \tilde{a}_{mi}\langle
i|\sigma _{3}|m\rangle +\tilde{b}_{mi}\langle m|\sigma _{3}|i\rangle \right)
\,.
\end{equation}
Combining this result with Eq. (\ref{I-E-Hartree}), we obtain the moment of
inertia
\begin{equation}
I=\frac{N_{c}}{4}\sum\limits_{i:\mathrm{occ}}\,\sum\limits_{m:\mathrm{nonocc}
}\left( \tilde{a}_{mi}\langle i|\sigma _{3}|m\rangle +\tilde{b}_{mi}\langle
m|\sigma _{3}|i\rangle \right) \,.  \label{I-coorect-app-0}
\end{equation}

The result (\ref{I-coorect-app-0}) for the moment of inertia $I$ coincides
with the well-known Thouless--Valatin formula \cite{TV-62} up to the
subtleties which differ our large-$N_{c}$ matrix $S$ (\ref{S-W}) from the
standard RPA matrix.

\section{$SU(6)$ Casimir operator}

\label{Casimir-appendix}

In this appendix we derive expression (\ref{C2-SU6-fermionic}) for the
$SU(6) $ Casimir operator. The $N^{2}-1$ generators $t^{a}$ of $SU(N)$ can be
normalized by the conditions 
\begin{equation}
\mathrm{Sp}(t^{a}t^{b})=\frac{1}{2}\delta ^{ab}\,,
\end{equation}
\begin{equation}
\mathrm{Sp\,}t^{a}=0\,,
\end{equation}
\begin{equation}
\sum\limits_{a=1}^{N^{2}-1}t_{AB}^{a}t_{CD}^{a}=\frac{1}{2}\left( \delta
_{AD}\delta _{BC}-\frac{1}{N}\delta _{AB}\delta _{CD}\right) \,.
\label{t-t-sum}
\end{equation}
In the case of $SU(6)$ the last relation results in
\begin{equation}
\sum\limits_{a=1}^{35}t_{AB}^{a}t_{CD}^{a}=\frac{1}{2}\left( \delta
_{AD}\delta _{BC}-\frac{1}{6}\delta _{AB}\delta _{CD}\right) \,.
\end{equation}
Therefore the fermionic realization of the $SU(6)$ Casimir operator
$C_{2}^{SU(6)}$ is
\[
C_{2}^{SU(6)}=\sum\limits_{a=1}^{35}\sum\limits_{A,B,C,D=1}^{6}\sum
\limits_{c,c^{\prime }=1}^{N_{c}}(\psi _{A,c^{\prime }}^{+}t_{AB}^{a}\psi
_{B,c^{\prime }})(\psi _{C,c}^{+}t_{CD}^{a}\psi _{D,c}) 
\]
\begin{equation}
=\sum\limits_{A,B=1}^{6}\sum\limits_{c,c^{\prime }=1}^{N_{c}}\left[ \frac{1}{
2}(\psi _{A,c^{\prime }}^{+}\psi _{B,c^{\prime }})(\psi _{B,c}^{+}\psi
_{A,c})-\frac{1}{12}(\psi _{A,c^{\prime }}^{+}\psi _{A,c^{\prime }})(\psi
_{B,c}^{+}\psi _{B,c})\right] \,.
\end{equation}
Returning to the $(f,s)$ form (\ref{A-fs}) of the $SU(6)$ multiindices, we
arrive at Eq. (\ref{C2-SU6-fermionic}).

\section{RPA equation in the toy model}

\label{RPA-toy-appendix}

In this appendix we describe the structure of the RPA equation for the
baryonic states of the toy model (\ref{H-toy-b}). In the Hartree picture for
these states we have one occupied nondegenerate level (\ref{eps-2}) and two
nonoccupied levels (\ref{eps-1}), (\ref{eps-3}) with degeneracies $3$ and $2$
respectively. This structure of the levels can be easily understood in terms
of the unbroken spin-flavor $SU(2)$ symmetry. The generators of this
symmetry are
\begin{equation}
G_{a}=J_{a}+T_{a}\quad (a=1,2,3)
\end{equation}
where $J_{a}$ is spin and $T_{a}$ is isospin. The occupied state has $G=0$,
whereas the nonoccupied levels have $G=1/2$ and $G=1$ with the corresponding
degeneracies $2G+1$. The RPA equation also can be split into two sectors
corresponding to $G=1/2$ and $G=1$. The nontrivial part of the RPA equation
corresponds to the $G=1$ sector. The RPA matrix $S$ can be computed using
the general expressions (\ref{S-W}), (\ref{W-def}), the explicit form of
$V_{A_{1}A_{2}|A_{3}A_{4}}$ in the toy model (\ref{V-b-Sigma-Phi}) and the
results for the single-particle levels (\ref{eps-1}), (\ref{eps-2}). In the
$G=1$ sector the RPA matrix $S$ looks as follows
\begin{equation}
S_{mini}=\frac{1}{2}\left( b_{12}+b_{21}\right) \delta _{mn}\left( 
\begin{array}{cc}
1 & 1 \\ 
1 & 1
\end{array}
\right) \quad (m,n:\quad G=1)\,.  \label{S-toy}
\end{equation}
The precise form of this matrix depends on the phase conventions for the
single-particle states. We fix the phases using the condition
\begin{equation}
\left( 
\begin{array}{c}
\langle m|\lambda _{a}|i\rangle \\ 
\langle i|\lambda _{a}|m\rangle
\end{array}
\right) =\delta _{am}\left( 
\begin{array}{c}
1 \\ 
1
\end{array}
\right) \quad (a,m=1,2,3)\,.
\end{equation}
Here we label the states $|m\rangle $ of the $G=1$ nonoccupied level with
numbers $m=1,2,3$.

Inserting the result for the matrix $S$ (\ref{S-toy}) into the general RPA
equation (\ref{RPA-compact}), we see that the RPA equation has only zero
eigenfrequencies $\Omega _{\nu }=0$ in agreement with the absence of the
harmonic excitations in the exact spectrum of the toy model.

\section{Rotational excitations in the toy model}

\label{I-calc-appendix}

In this appendix we compute the moment of inertia which determines the
spectrum of rotational excitations of baryonic states in the toy model
(\ref{H-toy-b}), using Eqs. (\ref{S-a-equation-0}), (\ref{I-coorect-app-0})
derived in Appendix \ref{Thouless-Valatin-appendix}.

In models, where the solution of Hartree equations has the spin-flavor
symmetry, the projector $P^{\mathrm{(occ)}}$ onto the occupied states commutes
with the generators of the combined spin-flavor rotations:
\begin{equation}
\lbrack P^{\mathrm{(occ)}},\left( \sigma _{a}+\tau _{a}\right) ]=0\,.
\end{equation}
Here the Pauli matrices $\sigma _{a},\tau _{a}$ act on spin and flavor
indices respectively. This symmetry allows us to obtain an equivalent
representation for the moment of inertia by replacing $\sigma
_{3}\rightarrow -\tau _{3}$ in Eqs. (\ref{S-a-equation-0}),
(\ref{I-coorect-app-0}). In the case of the $SU(3)$ flavor group we must use the
Gell-Mann matrix $\lambda _{3}$ instead of $\tau _{3}$. As a result we
arrive at the following representation for the moment of inertia
\begin{equation}
I=\frac{N_{c}}{4}\sum\limits_{i:\mathrm{occ}}\,\sum\limits_{m:\mathrm{nonocc}
}\left( a_{mi}\langle i|\lambda _{3}|m\rangle +b_{mi}\langle m|\lambda
_{3}|i\rangle \right) \,,  \label{I-coorect-app}
\end{equation}
where $a_{mi}$ and $b_{mi}$ are solutions of the equation
\begin{equation}
\sum\limits_{i^{\prime }:\mathrm{occ}}\,\sum\limits_{m^{\prime }:\mathrm{
nonocc}}S_{mim^{\prime }i^{\prime }}\left( 
\begin{array}{c}
a_{m^{\prime }i^{\prime }} \\ 
b_{m^{\prime }i^{\prime }}
\end{array}
\right) =\left( 
\begin{array}{c}
\langle m|\lambda _{3}|i\rangle \\ 
\langle i|\lambda _{3}|m\rangle
\end{array}
\right) \,,  \label{S-a-equation}
\end{equation}
and $S$ is the RPA matrix (\ref{S-RPA}).

We can rewrite Eq. (\ref{I-coorect-app}) in the form (\ref{I-correct}) but
one must keep in mind that matrix $S$ is degenerate.
Using expression~(\ref{S-toy}) for $S_{mim^{\prime }i^{\prime }}$ in the toy
model, we find from Eq. (\ref{S-a-equation})
\begin{equation}
\frac{1}{2}\left( b_{12}+b_{21}\right) \sum\limits_{m^{\prime }=1}^{3}\delta
_{mm^{\prime }}\left( 
\begin{array}{cc}
1 & 1 \\ 
1 & 1
\end{array}
\right) \left( 
\begin{array}{c}
a_{m^{\prime }i} \\ 
b_{m^{\prime }i}
\end{array}
\right) =\delta _{m3}\left( 
\begin{array}{c}
1 \\ 
1
\end{array}
\right) \,.
\end{equation}
The solution of this equation is
\begin{equation}
\left( 
\begin{array}{c}
a_{mi} \\ 
b_{mi}
\end{array}
\right) =\frac{1}{b_{12}+b_{21}}\delta _{m3}\left( 
\begin{array}{c}
1+c \\ 
1-c
\end{array}
\right) \,,  \label{a-b-solution}
\end{equation}
where $c$ is an uncertain constant. This constant cancels when we insert Eq.
(\ref{a-b-solution}) into Eq. (\ref{I-coorect-app}):
\[
I=\frac{N_{c}}{4}\,\sum\limits_{m=1}^{3}\left( a_{mi}\langle i|\lambda
_{3}|m\rangle +b_{mi}\langle m|\lambda _{3}|i\rangle \right) =\frac{N_{c}}{4}
(a_{mi}+b_{mi}) 
\]
\begin{equation}
=\frac{N_{c}}{2}\frac{1}{b_{12}+b_{21}}\,.  \label{I-correct-res}
\end{equation}

\section{Rigid-rotator Lagrangian}

\label{Rigid-rotator-appendix}

In this appendix we describe the structure of the effective rigid-rotator
Lagrangian for models containing colored fermions. Let $\pi _{0}$ be a
static solution of the Hartree equations (\ref{pi-epsilon}),
(\ref{pi-Hartree}). If $R$ is a transformation from the symmetry group $G$ of the
theory, then $R\pi _{0}R^{-1}$ is also a solution of Hartree equations. Now
let us apply a time-dependent transformation $R(t)$ to the static solution
$\pi _{0}$
\begin{equation}
\pi (t)=R(t)\pi _{0}R^{-1}(t)\,.
\end{equation}
Let us compute the bosonized action $S[\pi ]$ (\ref{S-bosonized}) for this
time-dependent configuration. The action (\ref{S-bosonized}) is nonlocal.
However, for slowly changing transformations $R(t)$ one can use the
``adiabatic approximation'' leading to the Lagrangian
\begin{equation}
L_{\mathrm{rot}}=\frac{1}{2}I_{ab}\Omega _{a}\Omega _{b}+w_{a}\Omega _{a}\,.
\label{L-rot-app}
\end{equation}
Here $\Omega _{a}$ is the angular velocity
\begin{equation}
\Omega =\Omega _{a}t_{a}=-iR^{-1}\dot{R}\,
\end{equation}
and $t_{a}$ are generators of the group $G$. The parameters of the
Lagrangian $L_{\mathrm{rot}}$ are given by equations
\begin{equation}
I_{ab}=N_{c}\sum\limits_{i:\,\mathrm{occ}}\,\,\sum\limits_{m:\,\mathrm{nonocc
}}\frac{\langle i|t_{a}|m\rangle \langle m|t_{b}|i\rangle +\langle
i|t_{b}|m\rangle \langle m|t_{a}|i\rangle }{\varepsilon _{m}-\varepsilon _{i}
}\,,  \label{I-ab-app}
\end{equation}
\begin{equation}
w_{a}=-N_{c}\sum\limits_{i:\,\mathrm{occ}}\langle i|t_{a}|i\rangle 
\label{w-a-app}
\end{equation}
where the summation runs over occupied ($i$) and nonoccupied ($m$)
eigenstates (\ref{pi-epsilon}) of the single-particle Hartree Hamiltonian.

It should be stressed that this ``derivation'' of the effective Lagrangian
(\ref{L-rot-app}) is inspired by the large-$N_{c}$ limit but not equivalent
to the systematic $1/N_{c}$-expansion.

\section{Model with $SU(2)$ flavor group}

\label{SU2-model-appendix}

The main part of this paper is devoted to the toy model (\ref{H-toy-b}) with
the flavor symmetry $SU(3)$. In this appendix we consider the $SU(2)$ analog
of this model. Let us take the same Hamiltonian (\ref{H-toy})
\[
H_{\mathrm{toy}}=\frac{1}{2N_{c}}\sum\limits_{f_{1},f_{2},f_{1}^{\prime
},f_{2}^{\prime }=1}^{2}\quad \sum\limits_{s_{1},s_{2},s_{1}^{\prime
},s_{2}^{\prime }=1}^{2}\left( \sum\limits_{c^{\prime }=1}^{N_{c}}\psi
_{f_{2}^{\prime }s_{2}^{\prime }c^{\prime }}^{+}\psi _{f_{1}^{\prime
}s_{1}^{\prime }c^{\prime }}\right) 
\]
\begin{equation}
\times V_{f_{2}^{\prime }s_{2}^{\prime }f_{1}^{\prime }s_{1}^{\prime
}|f_{2}s_{2}f_{1}s_{1}}\left( \sum\limits_{c=1}^{N_{c}}\psi
_{f_{2}s_{2}c}^{+}\psi _{f_{1}s_{1}c}\right)
\end{equation}
but now the indices $f_{k},f_{k}^{\prime }$ run only over the values $1,2$.
The coefficients $V_{f_{2}^{\prime }s_{2}^{\prime }f_{1}^{\prime
}s_{1}^{\prime }|f_{2}s_{2}f_{1}s_{1}}$ are determined by Eq.
(\ref{V-b-Sigma-Phi}). The diagonalization of this Hamiltonian is quite similar
to the case of the $SU(3)$ model. In the baryonic sector we have the states
with spin $J$ equal to isospin $T$
\begin{equation}
J=T=\frac{N_{c}}{2},\frac{N_{c}}{2}-1,\frac{N_{c}}{2}-2,\ldots
\end{equation}
The spectrum of energies of these states is
\begin{equation}
E=\frac{1}{2N_{c}}\left\{ b_{11}N_{c}^{2}+b_{22}N_{c}(N_{c}+3)+\left(
b_{12}+b_{21}\right) \left[ \frac{1}{2}N_{c}^{2}+2J(J+1)\right] \right\} \,.
\label{E-SU-2-exact}
\end{equation}
One can describe this spectrum in terms of the $1/N_{c}$-expansion. The
Hartree equations corresponding to the $1/N_{c}$-expansion make a subset of
equations that we had in the $SU(3)$ case. The single-particle energies
$\varepsilon _{1},\varepsilon _{2}$ are given by the same equations
(\ref{eps-1}), (\ref{eps-2}).

The leading $O(N_{c})$ part of the energy coincides with the $SU(3)$
expression (\ref{E1-res}) because it is determined by the same expression
for the Hartree energy (\ref{E-Hartree}). The $O(N_{c}^{-1})$ ``rotational'' 
$J(J+1)$ part of the energy also coincides with the $SU(3)$ case since it
comes from the RPA matrix $S$. The nontrivial part of the $SU(3)$ matrix $S$
coincides with the $SU(2)$ matrix $S$. Therefore the calculation of the
moment of inertia, performed in Appendix \ref{I-calc-appendix} for the
$SU(3) $ case, is also valid for $SU(2)$. The $O(N_{c}^{0})$ contribution to
the energy is given by equation (\ref{E0-general}) where we must use the
levels (\ref{eps-1}), (\ref{eps-2}): $\varepsilon _{i}=\varepsilon _{2}$,
$\varepsilon _{m}=\varepsilon _{1}$. Since the level $\varepsilon _{1}$ has
degeneracy 3, we obtain
\begin{equation}
E_{0}=-\frac{1}{2}\sum\limits_{m:\mathrm{nonocc}}\sum\limits_{i:\mathrm{occ}
}(\varepsilon _{m}-\varepsilon _{i})\,=-\frac{3}{2}(\varepsilon
_{1}-\varepsilon _{2})=\frac{3}{2}b_{22}\,,
\end{equation}
which agrees with the $O(N_{c}^{0})$ part of the exact solution
(\ref{E-SU-2-exact}).

\section{Extended monopole model}

\label{Extended-Guadagnini-model}

In this appendix we consider an extended version of the Guadagnini model
(\ref{L-monopole}) and show that in this extended model the picture of
``exotic'' states relying on the rigid-rotator approximation contradicts the
results based on the systematic semiclassical analysis corresponding to the
large-$N_{c}$ limit. We modify the model (\ref{L-monopole}) by adding a
second particle which interacts with the first one via a quadratic pair
potential:
\begin{equation}
L=\frac{\mu _{1}\mathbf{\dot{x}}_{1}^{2}+\mu _{2}\mathbf{\dot{x}}_{2}^{2}}{2}
-e(\mathbf{\dot{x}}_{1}\cdot \mathbf{A(x}_{1}\mathbf{)})-V(|\mathbf{x}
_{1}|)\,-\frac{1}{2}k\left( \mathbf{x}_{1}-\mathbf{x}_{2}\right) ^{2}.
\end{equation}
Note that this second particle does not interact with the magnetic field and
with the potential $V$.

Similar to Eqs. (\ref{mu-Nc}), (\ref{eg-B}), we assume the standard
semiclassical behavior inspired by the large-$N_{c}$ limit
\begin{equation}
\mu _{k},eg,V,k\sim N_{c}\,.
\end{equation}
Let us assume that potential $V$ has a minimum at some point $R\neq 0$:
\begin{equation}
V^{\prime }(R)=0\,.
\end{equation}
Then the static solutions of the equations of motion are described by the
conditions 
\begin{equation}
\mathbf{x}_{1}^{(0)}=\mathbf{x}_{2}^{(0)}\,,\quad |\mathbf{x}_{1}^{(0)}|=R\,.
\label{ext-mod-static-sol}
\end{equation}
Using the rotational invariance we can direct $\mathbf{x}_{1}^{(0)}=\mathbf{x
}_{2}^{(0)}$ along the third axis. Now let us write the linearized equations
of motion for small deviations from the static solution $\delta \mathbf{x}
_{k}(t)=\mathbf{x}_{k}(t)-\mathbf{x}_{k}^{(0)}$. The transverse part
(orthogonal to the third axis) does not mix with the longitudinal part. It
is convenient to use complex coordinates $z_{k}$ in the transverse plane
instead of $\delta \mathbf{x}_{k}^{\perp }$
\begin{equation}
\delta \mathbf{x}_{k}^{\perp }(t)\rightarrow z_{k}e^{i\omega t}\,.
\end{equation}
Then the linearized equations of motion take the form
\begin{equation}
-\mu _{1}\omega ^{2}z_{1}+h_{0}\omega z_{1}+k(z_{1}-z_{2})=0\,,
\label{ext-model-motion-1}
\end{equation}
\begin{equation}
-\mu _{2}\omega ^{2}z_{2}-k(z_{1}-z_{2})=0\,,  \label{ext-model-motion-2}
\end{equation}
where
\begin{equation}
h_{0}=\frac{eg}{R^{2}}\,.
\end{equation}
Equations of motion (\ref{ext-model-motion-1}), (\ref{ext-model-motion-2})
lead to the following equation for the frequency $\omega $
\begin{equation}
\det \left( 
\begin{array}{cc}
\mu _{1}\omega ^{2}-h_{0}\omega -k & k \\ 
k & \mu _{2}\omega ^{2}-k
\end{array}
\right) =0\,.
\end{equation}
Apart from the trivial solution $\omega =0$ one has three roots of the cubic
equation
\begin{equation}
\omega =k\left[ \left( \mu _{1}\omega -h_{0}\right) ^{-1}+\left( \mu
_{2}\omega \right) ^{-1}\right] \,.  \label{cubic-eq}
\end{equation}
These three roots determine the spectrum of the harmonic excitations.

Let us show that this spectrum of harmonic excitations differs from the
spectrum of the rigid-rotator Hamiltonian. Indeed, in the naive
rigid-rotator approximation the solution (\ref{ext-mod-static-sol}) leads by
analogy with Eq. (\ref{I-mu-r}) to the moment of inertia of the rigid
rotator 
\begin{equation}
I_{\mathrm{rot}}=\left( \mu _{1}+\mu _{2}\right) R^{2}\,.
\end{equation}
If one inserts this moment of inertia into the Hamiltonian
(\ref{H-rot-monopole}) and uses the quantization condition (\ref{J3-Witten}),
then similar to Eq. (\ref{E-exotic-monopole}) one arrives at the spectrum of
harmonic excitations
\begin{equation}
E_{\xi }=\frac{eg}{I_{\mathrm{rot}}}\left( \xi +\frac{1}{2}\right)
+O(1/N_{c})\,,\quad \xi =0,1,2,\ldots \,
\end{equation}
corresponding to the frequency
\begin{equation}
\omega _{\mathrm{rot}}=\frac{eg}{I_{\mathrm{rot}}}
=\frac{h_{0}}{\mu _{1}+\mu_{2}}\,.
\label{omega-rot-G2}
\end{equation}
But this frequency is different from the three roots of the cubic equation
(\ref{cubic-eq}). Only in the limit $k\rightarrow \infty $ (restoring the
rigidity of the system) one of the roots of equation (\ref{cubic-eq})
asymptotically approaches the frequency (\ref{omega-rot-G2}). But in a
general case the rigid-rotator approximation has nothing to do with the true
spectrum of harmonic excitations which is determined by equations
(\ref{det-K-0-0}), (\ref{K-delta-pi}).

\end{document}